%% file: main.tex
\documentclass[a4paper,USenglish]{article}
\usepackage{microtype}
\usepackage{amsfonts, amsmath, amssymb}
\usepackage{mathrsfs}
\usepackage{algorithm}
\usepackage{algorithmicx}
\usepackage{algpseudocode} 
\usepackage{afterpage}
\usepackage{graphicx}
\usepackage{datetime}
\usepackage{rotating}
\usepackage{hyperref}
\usepackage{pdflscape}
\usepackage{rotating}
\usepackage{placeins}
\newtheorem{theorem}{Theorem}[section]

\newtheorem{lemma}[theorem]{Lemma}

\newcommand{\qed}{\blacksquare}

\title{A  2/3-Approximation Algorithm for Vertex-Weighted Matching}

\author{Ahmed Al-Herz \\Computer Science Department, Purdue University, \\ West Lafayette IN 47907 USA, aalherz@purdue.edu 
\and 
Alex Pothen\\ Computer Science Department, Purdue University, \\ West Lafayette IN 47907 USA, apothen@purdue.edu
}

\begin{document}
\maketitle
\begin{abstract}
  \input{abstract.tex} 
\end{abstract}
\input{sectionIntroduction}
\input{sectionBackground} 
\input{sectionTwoThirdApprox}

\input{sectionCorrectness-TwoThirdsAlg}

\input{sectionExperiments}
\input{sectionConclusions}
\bibliography{paper}
\appendix
\input{Appendix}

\end{document}

%% file: abstract.tex
We consider the maximum vertex-weighted matching problem (MVM) for non-bipartite graphs, in which non-negative weights are assigned to the vertices of a graph, and we seek a matching that maximizes the sum of the weights of the matched vertices.  In earlier work we have described a $2/3$-approximation algorithm for the MVM on bipartite graphs~\cite{Dobrian+:VWM}. Here  we show that a $2/3$-approximation algorithm for MVM on non-bipartite graphs can be obtained by restricting the length of augmenting paths to at most three. The algorithm  has time complexity $O(m \log \Delta  + n \log n)$, where $n$ is the number of vertices, $m$ is the number of edges, and $\Delta$ is the maximum degree of a vertex. 

The approximation ratio of the algorithm is obtained by 
considering  {\em failed  vertices\/}, i.e., vertices that the approximation algorithm fails to match but the exact algorithm does. We show that there are two distinct heavier matched vertices that we can charge each  failed vertex to. Our proof techniques characterize the structure of augmenting paths in a novel way.

We have implemented the $2/3$-approximation algorithm and 
show that it runs in under a minute on graphs with tens of millions of vertices and hundreds of millions of edges. 
We compare its performance with five other algorithms:
an exact algorithm for MVM,  an exact algorithm for the maximum edge-weighted matching (MEM) problem, as well as three  approximation algorithms. The approximation algorithms include a $1/2$-approximation algorithm for MVM,  and $(2/3-\epsilon)$- and  $(1-\epsilon)$-approximation algorithms for the  MEM.  In our test set of nineteen problems, there are graphs on which the exact algorithms fail to terminate in $100$ hours. 
 In addition, the new $2/3$-approximation algorithm for MVM  outperforms the other approximation algorithms by either being faster (often by orders of magnitude)  or obtaining better weights. 
 

%% file: sectionIntroduction.tex
\section{Introduction}

We consider a variant of the matching problem in non-bipartite graphs in which weights are assigned to the vertices of a graph, the weight of a matching is the sum of the weights of the matched vertices, and we find a matching of maximum weight. We call this the maximum vertex-weighted matching problem (MVM). In this paper we describe a $2/3$-approximation algorithm for the MVM that has $O(m \log \Delta  + n \log n)$ time complexity, where 
$n$ is the number of vertices, $m$ is the number of edges, and $\Delta$ is the maximum degree of a vertex. We implement this algorithm as well as a few other approximation algorithms for this problem, and show that the $2/3$-approximation algorithm runs fast on large graphs, obtains  weights that are close to optimal,
and is faster or obtains greater weights than the other algorithms  on our test set. 

Consider an undirected vertex weighted graph $G=(V, E,\phi)$, where $|V| \equiv n$ is the number of vertices, $|E| \equiv m$ 
is the number of edges, and $\phi : V \mapsto R_{\geq 0}$ is a non-negative weight function on the vertices. The MVM problem can be solved in polynomial time by an exact algorithm~\cite{spencer1} with $O(\sqrt{n} m \log n)$ time complexity.
We have designed and implemented an exact algorithm with $O(mn)$ time complexity~\cite{Dobrian+:VWM}, since it is easier to implement, and it is well-known that the practical performance of a  matching algorithm does not necessarily correlate with its worst-case time complexity.
We show that this exact algorithm can be slow for many large graphs with millions of vertices and edges, and can even fail to terminate in $100$ hours. Thus, there is a  need for faster approximation algorithms that can return a matching with a guaranteed fraction of the maximum weight.

Many linear time approximation algorithms have been designed for the maximum edge weighted matching problem (MEM), but we are not aware of earlier approximation algorithms for the MVM problem on non-bipartite graphs. We have designed and implemented a $2/3$-approximation algorithm for MVM in bipartite graphs~\cite{Dobrian+:VWM}. 
 The MVM problem arises in applications  such as the design of network switches~\cite{tabatabaee1}, schedules for training of astronauts~\cite{bell1}, computation of sparse bases for the null space or the column space of a rectangular matrix~\cite{coleman3,pinar1,pothen3}, etc.

The  MVM problem can be transformed to a maximum edge weighted matching problem (MEM)  by assigning each edge a weight obtained by summing the weights at its endpoints.  Hence algorithms for the MEM  can be used to solve MVM problems. However, we have shown that this transformation can lead to increase in run times for an exact algorithm by three orders of magnitude or more~\cite{Dobrian+:VWM}. 
A simpler and more efficient exact algorithm is  obtained by solving the MVM problem directly by processing vertices in non-increasing order of weights and then matching an unmatched vertex to a heaviest unmatched neighbor it can reach by augmenting paths. In this sense the MVM problem is more similar to maximum cardinality matching than MEM. 
When we consider approximation algorithms,  restricting augmenting paths to length three does not lead to $2/3$-approximation algorithm for the MEM; however $2/3-\epsilon$-algorithms are available~\cite{pettie2004simpler}. 
The first $2/3$-approximation algorithm for MVM on {\em bipartite graphs\/} was proposed by us and our coauthors~\cite{Dobrian+:VWM}. The idea is to decompose the problem into two `one-side-weighted' problems, solve them individually by restricting the length of augmenting paths to at most three, and then combine the two matchings into a final matching by invoking the Mendelsohn-Dulmage  theorem~\cite{Mendelsohn+:theorem}. 
However, the proof technique used for bipartite graphs cannot be extended to non-bipartite graphs because the Mendelsohn-Dulmage theorem applies only to  the former. 
While the algorithm is simple to state, the proof that its approximation ratio  is $2/3$  requires  several new concepts and involves a careful study of the structure of augmenting paths.   

%% file: sectionBackground.tex
\section{Background and Related Work}
\label{sec:sectionBackground}

\subsection{Background on Matchings} 

We define the basic terms we use here, and additional background on matchings is available in several books, such as \cite{Schrijver:book}. 
The endpoints of an edge $(u,v)$ are the vertices $u$ and $v$. 
A  {\em matching\/} $M$ in a graph $G=(V,E)$ is a set of edges that do not share a common end-point; hence at most one edge from $M$ is incident on each vertex in the graph. 
An edge $(u,v) \in E$ 
is matched if $(u,v) \in M$, and  otherwise it is unmatched. A vertex $v \in V$ is matched if it is an end point 
of a matched edge,  and otherwise it is unmatched. 
A {\em heaviest unmatched neighbor\/} of $u$ is denoted by $HUN(u)$;  note that $HUN(u)$ might not be unique, but its weight is. 

A {\em path\/} $P$ in $G$ is a finite sequence of 
distinct vertices $\{v_i,v_{i+1},..v_{i+k}\}$ such that $(v_j,v_{j+1}) \in E$ for $i \leq j \leq i+k-1$. 
The length of a path $|P|$ is the number of edges in the path. A {\em cycle\/} is a path such that the first 
and the last vertex are the same. 
An {\em alternating path\/} $P$ with 
respect to $M$ is a path whose edges alternate between $M$-matched and $M$-unmatched edges. If the first and last vertices
on an $M$-alternating path $P$ are unmatched,  then it is an {\em $M$-augmenting path}, which necessarily has an odd number of edges.  The matching $M$ can be augmented  by
matching the edges in the  symmetric difference $M \oplus P$. 

When an augmentation is performed, we  
distinguish between the {\em origin\/} (the vertex from which an augmenting path search is initiated), and 
the {\em terminus\/} (the vertex at which the augmenting path search ends). We will denote the origin of the $i-$th augmentation step  by $o_i$, and the terminus by $t_i$,  where $i \geq 1$. 
When a vertex is not explicitly denoted by $o_i$ or $t_i$, then it could be either an origin or a terminus, or neither,  unless mentioned otherwise. 
Note that if we begin with the empty matching, then  for each matched edge we need one augmentation step,  so that $|M|$ is equal to  the number of origins or termini.
We will say that the $i$-th origin and the $i$-th terminus {\em correspond\/} to each other, so that $o_i (t_i)$ is the corresponding vertex of the vertex $t_i (o_i)$. 
An {\em alternating path\/} $P$ with respect to two matchings, $M_1 \oplus M_2$, is a path whose edges alternate between $M_1$-matched edges and $M_2$-matched edges.

\subsection{Related Work}

We now describe more fully the work we have done  earlier  with our colleagues  on exact algorithms for MEM  on general graphs, and a $2/3$-approximation algorithm for this problem on bipartite graphs~\cite{Dobrian+:VWM}. 
When vertex weights are non-negative, we can choose a maximum vertex-weighted matching to be one of maximum cardinality, and from now on we assume that this choice has been made. 

For a matching $M$, an {\em $M$-reversing path \/} is an alternating path with even number of edges consisting of an equal number of matched and unmatched edges. 
An {\em $M$-increasing path\/} is an $M$-reversing path whose unmatched endpoint has higher weight than its matched endpoint. 
By switching the matched and unmatched edges on this path, we can increase the weight of the matching, much as we would  using an augmenting path. 
There are two ways of characterizing maximum vertex-weighted matchings in a general graph. The first is that a matching $M$ is an MVM if and only if there is $(i)$ neither an $M$-augmenting path  ($(ii)$ nor an  $M$-increasing path in the graph.
The second is to list the weights of matched vertices in non-increasing order in a vector (this is the weight vector of  the matching $M$). 
Then a matching $M$ is an MVM if and only if its weight vector is 
lexicographically maximum among all the weight vectors of  matchings. 

These two characterizations lead to two extreme algorithms for computing 
an MVM. The first begins with the empty matching, and at each step matches a currently heaviest unmatched vertex to a heaviest unmatched vertex it can reach by augmenting path. In this algorithm once a vertex is matched, it will always remain matched, since augmentation does not change a matched vertex to an unmatched vertex. 

An algorithm with the approximation ratio of $2/3$ for MVM on bipartite graphs was designed and implemented in \cite{Dobrian+:VWM}. 

The second exact algorithm for solving the MVM problem begins with a maximum cardinality matching. It then looks for increasing paths or cycles with respect to the current matching and terminates when there are none such. This second, speculative,  algorithm has the advantage that it has more concurrency whereas the first algorithm has to process vertices in a specified order. 
We will discuss the speculative algorithm in future work. 

Now we turn to  approximation algorithms that have been designed for the maximum edge weighted matching problem (MEM).
The well-known Greedy algorithm~\cite{avis1983survey} iteratively adds a  heaviest edge to the matching,  and deletes all edges incident on the endpoints of the added edge.
 This algorithm is $1/2$-approximate and requires $O(m \log n)$ time. 
Another $1/2$-approximation algorithm, the Locally Dominant edge algorithm~\cite{preis1999linear},  avoids sorting the edges by choosing  locally dominant edges (an edge that is the  heaviest edge incident on both of its  endpoints) to add to the matching. 
A more recent $1/2$-approximation algorithm is 
 the Suitor algorithm~\cite{manne2014new},  which employs a proposal-based approach similar to the classical algorithms for stable matching. The Suitor and other  algorithms have been extended to find $1/2$-approximate $b$-Matchings~\cite{Khan1}. 
 Other papers improve the performance ratios:  For any fixed $\epsilon > 0$, $(2/3-\epsilon)$- and $(3/4-\epsilon)$-approximation algorithms have been proposed~\cite{drake2003improved, DuanP10b, Hanke2010, pettie2004simpler, Maue+:matching}.  
 Furthermore, a $(1-\epsilon)$-approximation algorithm, 
 based on a scaling approach has been proposed by Duan and Pettie~\cite{DuanP-approxMWM} which has time complexity $ O(m \, \epsilon^{-1} \log(\epsilon^{-1}))$.  We show that the $(1-\epsilon)$-approximation algorithm  when applied to the MVM problem is significantly slower than the $1/2$- and $2/3$-approximation algorithms, and surprisingly, does not compute greater matching weights
 for relevant values of $\epsilon$. 
 The MEM problem appears in applications such as placing large elements on the diagonal of sparse matrices~\cite{duff1, duff2}, multilevel graph partitioning~\cite{karypis1}, scheduling, etc. 
 

%% file: sectionTwoThirdApprox.tex
\section{A Two-third Approximation Algorithm for MVM}
\label{sec:sectionTwoThirdApprox}
In this section we describe a $2/3$-approximation algorithm for MVM, discuss its relation to an exact algorithm, and then  prove its correctness. 

\subsection{Exact and  $2/3$-Approximation Algorithms}
The approximation algorithm, described in Algorithm~\ref{Approx}, sorts the vertices in non-increasing order of weights, and inserts the sorted vertices  into a queue $Q$. The algorithm begins with the empty matching, and attempts to match the vertices in $Q$ in the given order. Each unmatched vertex $u$ is removed from $Q$, and beginning at $u$ the algorithm searches for a heaviest unmatched vertex $v$ reachable by an augmenting path of length at most {\em three}.
 If such an augmenting path is found, then the matching is augmented by the path that leads to a heaviest unmatched vertex,  and the vertex $v$ is also removed from $Q$. If no augmenting path of length at most three is found, we search from the next heaviest unmatched vertex (even though longer augmenting paths might exist in the graph). 
 The algorithm terminates when all vertices are processed. 
 
 The $2/3$-approximation algorithm may be viewed as one obtained
 from an exact algorithm for MVM. In the exact  algorithm for MVM,
 at each step we search from a currently heaviest unmatched vertex for a heaviest unmatched vertex reachable by an augmenting path of any length.
 If an augmenting path is found, we choose the path that leads to a heaviest unmatched vertex, and then augment by this path. 
 If no augmenting path is found, we search from the next heaviest unmatched vertex. 
 This algorithm was proved correct by Dobrian, Halappanavar, 
 Pothen and Al-Herz in \cite{Dobrian+:VWM}. The time complexity of this 
 algorithm is $O(nm)$. 

 Consider running the Exact algorithm and the 2/3-approximation algorithm simultaneously using the vertices in the same queue $Q$. Both consider vertices in non-increasing order of weights, and break ties among weights consistently.  If a vertex $u$ is matched by the exact algorithm but not by the approximation algorithm (because the augmenting path is longer than three), then we call $u$ a {\em failure\/} or a failed vertex, because the approximation algorithm failed to match it while the exact algorithm succeeded.

\begin{algorithm}
\caption{Input: A graph $G$ with weights $\phi$ on the vertices. Output: A matching $M$. Effect: Computes a 2/3-approximation to a maximum vertex-weighted matching.}
\label{Approx}
\begin{algorithmic}[1]
\Procedure{TWOTHIRD-APPROX($G=(V,E,\phi)$)}{}
\State $M \gets \emptyset$;
\State$Q 	\gets V$;
\While{$Q \neq \emptyset$ }
\State $u \gets heaviest(Q)$;
\State $Q \gets Q - u$;
\State{Let $v$ denote a heaviest unmatched vertex reachable from $u$ by an \indent \indent augmenting  path $P$ of length at most three;}
\If{$P$ is found}
\State $M \gets M \oplus P$; \ $Q \gets Q - v$;
\EndIf
\EndWhile
\EndProcedure
\end{algorithmic}
\end{algorithm}


\subsection{Time Complexity of the Two-Thirds Approximation Algorithm}
\begin{theorem} 
The time complexity of the Two-Thirds approximation algorithm is $O(m \log \Delta + n \log n)$, where $\Delta$ is the maximum degree.
\end{theorem}

{\bf Proof}:
We sort the adjacency list of each vertex in non-increasing 
order of weights, and maintain a pointer to a heaviest unmatched neighbor of each vertex. Since the adjacency list 
is sorted, each list is searched once from highest  to lowest
weight in the algorithm. 

Let $N(u)$ be the set of neighbors of a vertex $u$ and $d(u) = |N(u)|$. In each iteration of the while loop, we choose an unmatched vertex $u$ and examine all vertices in $N(u)$ to
find a heaviest unmatched neighbor, if one exists. 
If $u$ has a matched neighbor $v$,  then we form an augmenting path of length three by taking the matched edge $(v,w)$, and 
finding a heaviest unmatched neighbor $x$ of $w$. 
All neighbors of $u$, unmatched and matched, can be found 
in $O(d(u))$ time, and finding the matched vertex $w$ and a 
heaviest unmatched neighbor $x$ can be done in constant time,
since the adjacency lists are sorted. Thus the search for 
augmenting paths in the algorithm takes $O(m)$ time. 
Sorting the adjacency lists takes time proportional to 
$$\sum_u d(u) \log d(u) \leq \sum_u d(u) \log \Delta = m \log \Delta.$$
Sorting the vertices in non-increasing order of weights takes $O(n \log n)$ time. 
$\quad \qed$

%% file: sectionCorrectness-TwoThirdsAlg.tex
\subsection{Correctness of the Algorithm}
\label{subsec:sectionApproximationQuality}

In this subsection we will prove (Theorem~\ref{thm:mainTheorem}) that Algorithm~\ref{Approx} computes a $\frac{2}{3}$-approximate MVM, $M_A$. Let $\phi(F)$ denote the sum of the weights of the failures,  $\phi(M_A)$ the weight of the approximate matching, and $\phi(M_{opt})$ the weight of an optimal matching. In order to prove the theorem, it suffices to prove  that $\phi(F) \leq \frac{1}{2}\phi(M_A)$, since $\phi(M_{opt}) \leq \phi(M_A) + \phi(F)$.

To prove that $\phi(F) \leq \frac{1}{2}\phi(M_A)$,  we show that for every failure there are two distinct vertices that are  matched in $M_A$, with weight at least as heavy as the failure. This is achieved in Lemma~\ref{lem:compensating-vertices} by considering $M_{opt} \oplus M_A$-alternating paths, using a charging technique in which each failure charges two distinct vertices matched in $M_A$. Each failure is an endpoint of the $M_{opt} \oplus M_A$-alternating path. The two distinct vertices are obtained as the {\em corresponding vertices\/} (the other ends of the augmenting paths) of two of the first three vertices on the $M_{opt} \oplus M_A$-alternating path.

We prove the approximation ratio by means of several Lemmas. The key Lemma~\ref{lem:compensating-vertices}  is proved using Lemmas~\ref{lem:hun},~\ref{lem:v1},~\ref{lem:v3} and ~\ref{lem:noChange}. We begin by proving each of the latter Lemmas. 

\begin{lemma}
\label{lem:hun}
Let $(u,v)$ be a matched edge in a matching $M$ at some step in the $2/3$-approximation algorithm, and let $w = HUN(v)$ be 
a heaviest unmatched neighbor of $v$. 
Suppose $(u,v)$ is changed to 
a matched edge $(u,v')$ in a future augmentation step, and  
let $w' = HUN(v')$ denote a heaviest unmatched neighbor of $v'$,  then $\phi(w) \geq \phi(w')$.
\end{lemma}

{\bf Proof}:
The proof is by induction on $i$, the number of augmentation steps that include $u$ on the augmenting path. 
Let $v'_i$ be the matched neighbor of $u$ after $i$ augmentation steps involving $u$, and let $w'_i$ be its heaviest unmatched neighbor $HUN(v'_i)$. 
There are two possible augmentation steps that include 
the matched edge $(u,v)$. 
(1) $\{o_i, u, v, t_i \}$, and (2)  $\{o_i, v, u, t_i \}$,
where $o_i$ ($t_i$) is the origin (terminus) of the augmenting path.

\begin{figure}[!thp]
\centering
\includegraphics[scale=0.5]{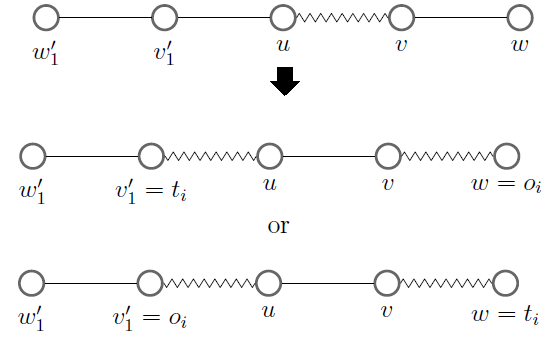}
\caption{Lemma~\ref{lem:hun}: Base case.}
\label{Fig:Lemma1BaseCase}
\end{figure}

For the base case, $i=1$), consider  Figure~\ref{Fig:Lemma1BaseCase}.  If  the augmentation path is $\{o_1 = w, v, u, t_1 = v'_1 \}$, clearly $\phi(w) \geq \phi(w'_1)$, since the algorithm processes vertices in non-increasing order of weights.  If the augmenting path is  $\{o_1 = v'_1, u, v, t_1 = w \}$, then $\phi(w) \geq \phi(w'_1)$ because $w$ was matched in preference to $w'_1$. 

Assume the claim is true for $k$ augmentation steps. By using the same argument as in the base case we have $\phi(w'_k) \geq \phi(w'_{k+1})$ at the $k+1$-st augmentation step. Now by the inductive hypothesis we have $\phi(w) \geq \phi(w'_k)$, and 
 by combining the two inequalities, we obtain $\phi(w) \geq \phi(w'_{k+1})$.
$\quad \qed$

\begin{lemma}
\label{lem:v1}
 Let $M_A^x$ denote the $2/3$-approximate matching at the $x^{\text{th}}$ failure $f_x$,  and let $P = \{f_x, v_1, v_2, ...\}$ be an alternating path that begins with $f_x$ in $M_{opt} \oplus M_A^x$.
 \newline (1) If $v_1$ is an origin $o_i$ of some prior augmentation step, then $\phi(t_i) \geq \phi(f_x)$.
 \newline (2) $\phi(v_2) \geq \phi(f_x)$.
\end{lemma}

{\bf Proof}:
(1) If $v_1$ is an origin $o_i$,  then we have $\phi(t_i) \geq \phi(f_x)$, because $t_i$ was matched in preference to $f_x$.

\begin{figure}[!thp]
\centering
\includegraphics[scale=0.5]{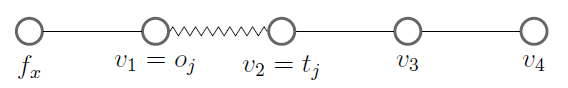}
\caption{ Lemma~\ref{lem:v1} Case (b): $v_2$ is a terminus that is matched by an augmenting path that includes $v_1$.}
\label{Fig:Remark1-2}
\end{figure}

\begin{figure}[!thp]
\centering
\includegraphics[scale=0.5]{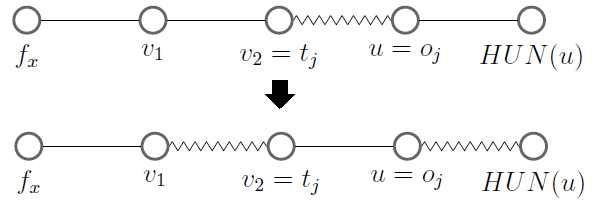}
\caption{Lemma~\ref{lem:v1} Case (c): $v_2$ is a terminus that is matched by an augmenting path that includes $u \neq v_1$. }
\label{Fig:Remark1-3}
\end{figure}

(2) In this case, we have to consider three possibilities.
\newline 
(a) The vertex $v_2$  is an origin, in which case $\phi(v_2) \geq \phi(f_x)$, since $v_2$ was processed before $f_x$. 
\newline
(b) The vertex $v_2$ is a terminus that is matched by an augmenting path that includes $v_1$. An example of this case
 is shown in Figure~\ref{Fig:Remark1-2}. In this case we have two possibilities: either $v_1$ is an origin and $v_2$ is the corresponding terminus, or $v_1$ is previously matched in which case we have an augmenting path $\{o_i, x, v_1, v_2\}$. In both possibilities  $v_2$ was matched in preference to $f_x$, so $\phi(v_2) \geq \phi(f_x)$.
 \newline 
(c) The vertex $v_2$ is a terminus that is matched by an augmenting path that includes a vertex $u \neq v_1$,
where $u$ is adjacent to $v_2$. An example of this case is shown in Figure~\ref{Fig:Remark1-3}. Let $HUN(u)$ be a heaviest unmatched neighbor of $u$ after $v_2$ is matched. In this case, again we have two possibilities:  $u$ is an origin and $v_2$ is the corresponding terminus, or $u$ is previously matched in which case we have an augmenting path $\{o_i, x, u, v_2\}$. In both possibilities $v_2$ was matched in preference to $HUN(u)$ so we have $\phi(v_2) \geq \phi(HUN(u))$. By Lemma~\ref{lem:hun} when the matched edge $(u,v_2)$ is changed to the matched edge $(v_1,v_2)$, we have $\phi(HUN(u)) \geq \phi(f_x)$.   
By combining these two inequalities, we obtain 
$\phi(v_2) \geq \phi(f_x)$. 
$\quad \qed$

\begin{lemma}
\label{lem:v3}
Let $M_A^x$ denote the $2/3$-approximate matching at the $x^{\text{th}}$ failure $f_x$, and let $P = \{f_x, v_1, v_2, v_3, ...\}$ be an $M_{opt} \oplus M_A^x$-alternating path that begins with $f_x$. If the vertex $v_3$ is an origin $o_i$ of some prior augmentation step in the Approximation algorithm, and if $\phi(t_i) < \phi(f_x)$,  then 
 1) immediately prior to the step when the Approximation algorithm matches the vertex $v_3$, the vertex $v_2$ is matched to a vertex  $u \neq v_1$,  and $\{v_2, v_3, u\}$ is a cycle.
 \newline
 2) the $i$-th augmenting path is $\{v_3=o_i,u,v_2,t_i\}$.
\end{lemma}

{\bf Proof}:
1) First we will establish that $v_2$ is matched to some vertex $u$ prior to the step when $v_3$ is matched. To obtain a contradiction, assume that $v_2$ is not matched to some vertex $u$ prior to the step of matching $v_3$. Then after $v_3$ is matched, the terminus $t_i$ is either $v_2$ or a vertex that is matched in preference to $v_2$. In both possibilities we have 
$\phi(t_i) \geq \phi(v_2)$. 
We know from Lemma~\ref{lem:v1} that 
$\phi(v_2) \geq \phi(f_x)$. 
Combining the two inequalities,  we have 
$\phi(t_i) \geq \phi(f_x)$, which contradicts the assumption in the Lemma. 

Now we show that the vertex $u \neq v_1$. Assume for a contradiction  that $u = v_1$, then at the step of matching $v_3$ there exists an augmenting path from $v_3$ to $f_x$ of length three. After we match $v_3$, we have $\phi(t_i) \geq \phi(f_x)$, since it was matched in preference to $f_x$. This again contradicts the assumption in the Lemma. 

\begin{figure}[!thp]
\centering
\includegraphics[scale=0.5]{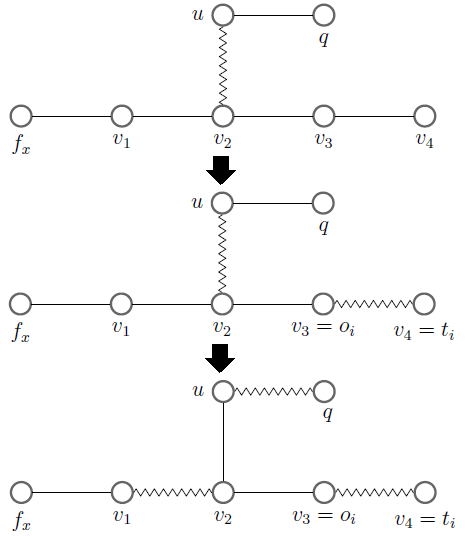}
\caption{Lemma~\ref{lem:v3}: The case where $v_3 \neq HUN(u)$.}
\label{Fig:Lemma4-1}
\end{figure}

Now we show that $\{v_2, v_3, u\}$ is a cycle by showing that 
$v_3 = HUN(u)$. Assume $v_3 \neq HUN(u)$ and let some vertex $q=HUN(u)$,  as shown in Figure~\ref{Fig:Lemma4-1}. Note that by Lemma~\ref{lem:hun} we have $\phi(q) \geq \phi(HUN(v_1)) \geq \phi(f_x)$ (A),  since we know the 
matching edge $(v_2,u)$ is changed to $(v_2,v_1)$. Also, 
immediately prior to the step when $v_3$ is matched, 
there exists an augmenting path of length three from $v_3$ to $q$. So after we match $v_3$, $t_i$ is either $q$ or a vertex that is matched in preference to $q$, so $\phi(t_i) \geq \phi(q)$ (B). 
Combining (A) and (B) we get $\phi(t_i) \geq \phi(f_x)$. Thus, $v_3 = HUN(u)$. Hence $\{v_2, v_3, u\}$ is a cycle since we have established the existence of the edge $(u,v_3)$ (the existence of the other two edges of the cycle were established earlier).\newline

\begin{figure}[!thp]
\centering
\includegraphics[scale=0.5]{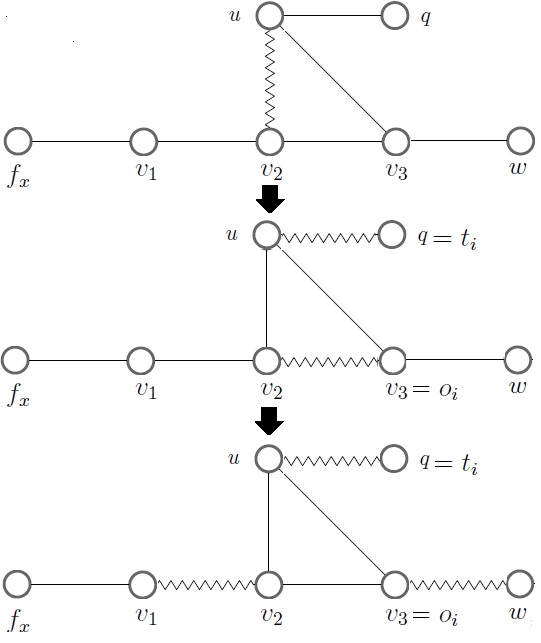}
\caption{Lemma~\ref{lem:v3}, (2) Case 1: The augmentation step is $\{o_i,v_2,u,t_i\}$.}
\label{Fig:Lemma4-2}
\end{figure}

2) We establish this result by contradiction as well. 
Suppose the augmenting path is not $\{o_i,u,v_2,t_i\}$. Then we have two cases:\newline
Case 1: The augmenting path is $\{o_i,v_2,u,t_i\}$ as shown in Figure~\ref{Fig:Lemma4-2}. 
In this case there must exist an unmatched vertex $w$ adjacent to $v_3$, since after matching the edge $(v_2,v_3)$ it must be changed to $(v_2,v_1)$ by an augmenting path of length three. After matching $v_3$, assume without loss of generality that $w$ becomes $HUN(v_3)$. After the augmentation step, we have $\phi(t_i) \geq \phi(w)$ (A), since there existed an augmenting path from $v_3$ to $w$ when $t_i$ was matched. Also, $(v_2,v_3)$ was matched in this step, and it must be changed to the matched edge $(v_2,v_1)$. By Lemma~\ref{lem:hun} we have $\phi(w=HUN(v_3)) \geq \phi(HUN(v_1)) \geq \phi(f_x)$ (B). Combining (A) and (B), we obtain $\phi(t_i) \geq \phi(f_x)$.  Again we have a contradiction of the condition of the Lemma. 

\begin{figure}[!thp]
\centering
\includegraphics[scale=0.5]{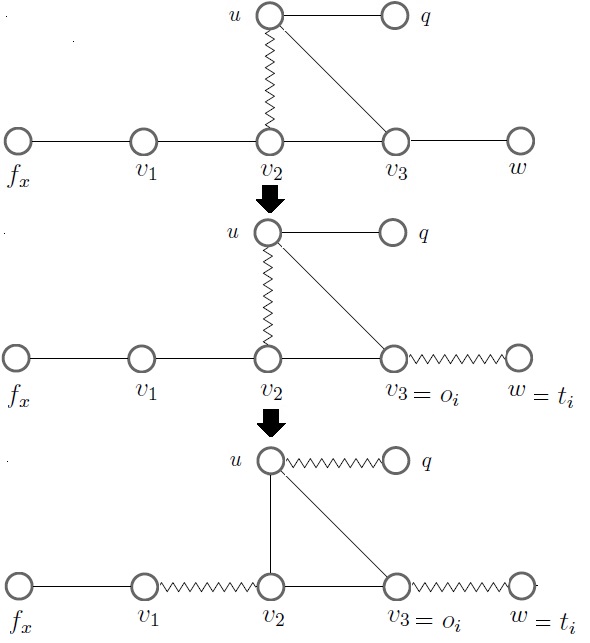}
\caption{Lemma~\ref{lem:v3} (2) Case 2: the augmentation step does not include the edge $(v_2,u)$.}
\label{Fig:Lemma4-3}
\end{figure}

Case 2: The augmentation step does not include the edge $(v_2,u)$ as shown in Figure~\ref{Fig:Lemma4-3}. 
In this case there must exist an unmatched vertex $q$ adjacent to $u$ since the matched edge $(v_2,u)$ must be changed to $(v_2,v_1)$ by an augmenting path of length three. After matching $v_3$, assume without loss of generality that $q$ becomes $HUN(u)$. After the augmentation step, we have $\phi(t_i) \geq \phi(q)$ (A), since there existed an augmenting path from $v_3$ to $q$. Note that $(v_2,u)$ is still matched and must be changed to $(v_2,v_1)$. By Lemma~\ref{lem:hun} $\phi(q=HUN(u)) \geq \phi(HUN(v_1)) \geq \phi(f_x)$ (B). Again, combining (A) and (B), we obtain $\phi(t_i) \geq \phi(f_x)$.

In both cases we obtain $\phi(t_i) \geq \phi(f_x)$, 
a contradiction to the condition of the Lemma. Therefore, the $i$-th augmentation step must be $\{v_3=o_i,u,v_2,t_i\}$.
$\quad \qed$

\begin{lemma}
\label{lem:noChange}
Consider the symmetric difference $M_{opt} \oplus M_A^x$, corresponding to the $2/3$-approximate matching at the $x$-th failure. 
Let $P = \{f_x, v_1, v_2, v_3, v_4\}$ be an $M_{opt} \oplus M_A^x$-alternating path, then the alternating subpath  $P = \{f_x, v_1, v_2, v_3\}$ will not change in future augmentation steps of  the approximation algorithm.
\end{lemma}

\begin{figure}[!thp]
\centering
\includegraphics[scale=0.5]{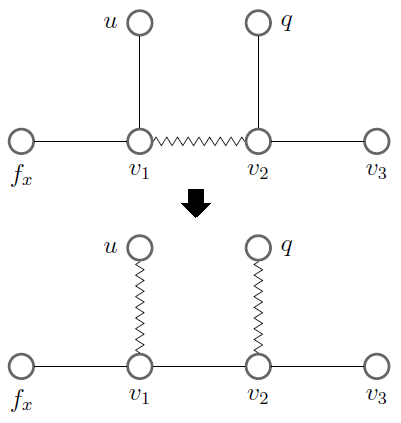}
\caption{Lemma~\ref{lem:noChange}: Augmenting the path $\{u,v_1,v_2,q\}$ after $f_x$ is determined to be a failure.}
\label{Fig:Lemma5}
\end{figure}

{\bf Proof}:
 Assume for the sake of contradiction that after $f_x$ is determined to be a failure, the edge $(v_1,v_2)$ is changed by a future augmenting path of length three,  say $\{u,v_1,v_2,q\}$,  as shown in Figure~\ref{Fig:Lemma5}. Then, the augmenting path $\{f_x,v_1,v_2,q\}$ must exist
 when $f_x$ was determined as a failure, and in this case $f_x$ could not have been a failure.  Hence the matched edge $(v_1,v_2)$ in the approximate matching $M_A^x$ cannot be changed in future augmentations.
$\quad \qed$

\begin{lemma}
\label{lem:compensating-vertices}
Consider the symmetric difference $M_{opt} \oplus M_A$,
where $M_A$ is the matching computed by the $2/3$-Approximation algorithm. For every failure $f$ there are two distinct matched vertices in $M_A$ that are at least as heavy as $f$.
\end{lemma}
{\bf Proof}:
First run the approximation algorithm and at the  $i$-th augmentation step  label the origin by $o_i$ and the terminus by $t_i$. Recall that we denote $o_i$ as the corresponding vertex of $t_i$,  and vice versa.
 Consider the symmetric difference between $M_{opt}$ and $M_A$ which results in alternating paths and cycles. 
 We can ignore alternating cycles since every vertex in a cycle is matched in both $M_{opt}$ and $M_A$. 
Since failures are matched by the optimal matching but not the approximate matching, they are at the ends of alternating paths. 

 By Lemma~\ref{lem:noChange} the first four vertices of an alternating path beginning with a failure  do not change,  which makes it possible to identify the origins and termini which are used to construct the alternating path. 
 We will number each failure $f_x$ in the order that it was discovered in the approximation algorithm. A failure $f_x$ could be an end of an alternating path which has  one failure or two failures. We will consider these two types of alternating paths in the following.  
 
\begin{figure}[!thp]
\centering
\includegraphics[scale=0.5]{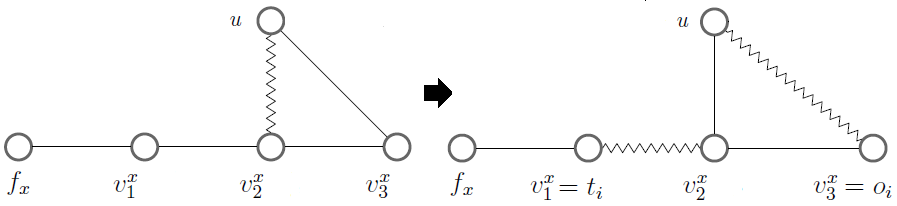}
\caption{Lemma~\ref{lem:compensating-vertices}, $i-2$: The corresponding terminus is strictly lighter than the failure $f_x$.}
\label{Fig:lemma6-1}
\end{figure}

 $(i)$ First consider an alternating path with one failure, and denote the path as $P = \{f_x, v_1^x, v_2^x, v_3^x, v_4^x\}$.  We charge two distinct vertices for $f_x$ as follows: 
 \newline 
 $(i-1)$ If the vertex $v_1^x$ is a terminus,  then charge the corresponding origin,  which must be at least as heavy as the failure $f_x$ since it was processed before $f_x$.  If $v_1^x$ is an origin then charge the corresponding terminus,  which by Lemma~\ref{lem:v1}~(1) must be at least as heavy as $f_x$. 
 \newline
 $(i-2)$ If the vertex $v_3^x$ is a terminus,  then charge 
 the corresponding origin which must be at least as heavy as the failure $f_x$ since it was processed before $f_x$. 
 If $v_3^x$ is an origin, and the corresponding terminus is at least as heavy as $f_x$, then charge the corresponding terminus. If the corresponding terminus is strictly lighter than $f_x$, then by Lemma~\ref{lem:v3} we have 
 immediately prior to the step in which $v_3^x$ is matched, the vertex $v_2^x$ is matched to some vertex $u$, $u \neq v_1^x$ such that  $\{v_2^x, v_3^x, u\}$ is a cycle,  as shown in Figure~\ref{Fig:lemma6-1}. In this case we consider $v_2^x$ instead of $v_3^x$ to find a vertex to charge.  If the vertex $v_2^x$ is a terminus (in a prior augmentation step),  then charge the corresponding origin which must be at least as heavy as $f_x$, since it was processed before the latter. If $v_2^x$ is an origin in the prior augmentation step,  then charge the corresponding terminus which must be at least as heavy as $f_x$ since it was matched in preference to $v_3^x$ which is an origin. \newline

 $(ii)$ Now we consider an alternating path with two failures $f_x$ and $f_y$ as its endpoints. 
 We assume without loss of generality that $\phi(f_x) \geq \phi(f_y)$.\newline 
For the failure $f_x$ we charge two distinct vertices as we did in Part $(i)$ of this Lemma. Now we consider charging for the failure $f_y$.  If the length of the alternating path is at least seven edges,  then we can label two alternating subpaths $\{f_x, v_1^x, v_2^x, v_3^x\}$ and $\{f_y, v_1^y, v_2^y, v_3^y\}$, and these do not overlap. 
Hence we can charge two distinct vertices for $f_y$ as we did in Part $(i)$ of the Lemma. 

If the length of the alternating path is five then $\{ v_2^x, v_3^x\}$ and $\{ v_2^y, v_3^y\}$ overlap. Thus $v_2^x = v_3^y$, and $v_3^x = v_2^y$. So, we charge one vertex $v_1^y$ for $f_y$ as we did in $(i-1)$ and we will charge the other distinct vertex as follows. 
\newline 
Case 1: If $f_x$ charged the corresponding vertex of $v_2^x$ then $f_y$ must charge the corresponding vertex of $v_2^y = v_3^x$. Referring to $(i-2)$, the vertex $f_x$ charged the corresponding vertex of $v_2^x$ because $v_3^x = v_2^y$ must be an origin and the corresponding terminus is strictly lighter than $f_x$. Let the  origin $v_3^x$ be denoted by $o_i$, and the corresponding terminus be $t_i$, for some augmentation step $i$. By Lemma ~\ref{lem:v3} we have (1) at the step of matching $v_3^x$ but before it is matched, $v_2^x$ is matched to some $u$, where $u \neq v_1^x$, and $ \{v_2^x, v_3^x, u\}$ is a cycle;  (2) the augmenting path is $\{v_3^x=o_i, u, v_2^x, t_i\}$. 

\begin{figure}[!thp]
\centering
\includegraphics[scale=0.5]{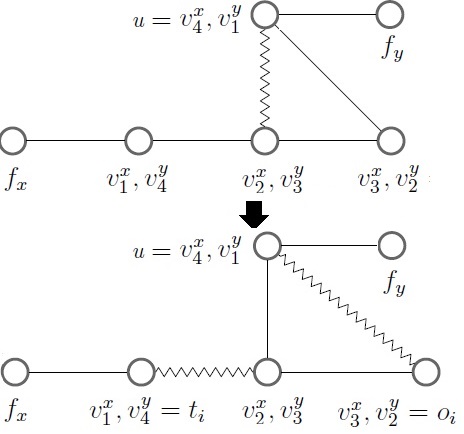}
\caption{Lemma~\ref{lem:compensating-vertices}, Case 1, Subcase 1:  The failure $f_y$ is adjacent to $u$. }
\label{Fig:case1}
\end{figure}

We will show that $ \phi(t_i) \geq \phi(f_y)$, and thus $f_y$ can be charged to $t_i$. We consider two subcases:\newline
Subcase 1: $f_y$ is adjacent to $u$, as shown in Figure~\ref{Fig:case1}. Note that $\phi(t_i) \geq \phi(f_y)$, since at the step of matching $v_3^x$ there existed an augmenting path from $v_3^x$ to $f_y$. 

\begin{figure}[!thp]
\centering
\includegraphics[scale=0.5]{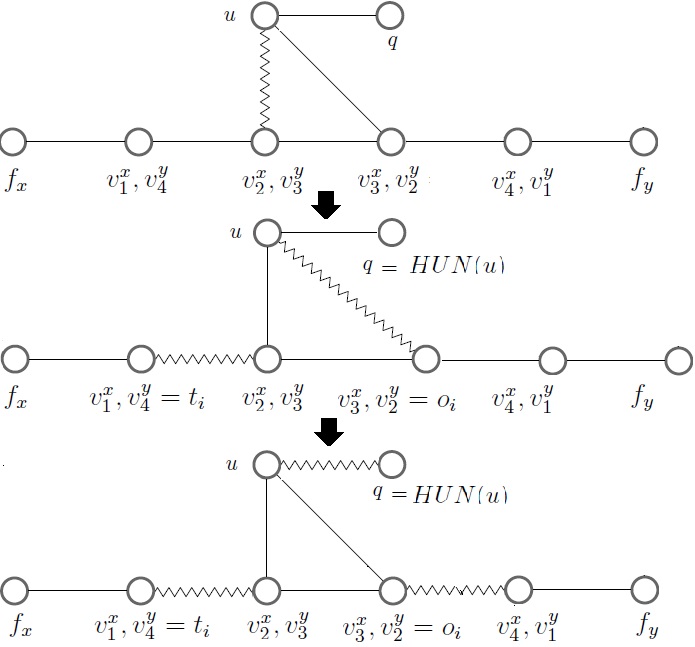}
\caption{Lemma~\ref{lem:compensating-vertices}, Case 1, Subcase 2: The failure $f_y$ is not adjacent to $u$. }
\label{Fig:case2}
\end{figure}

Subcase 2: The failure $f_y$ is not adjacent to $u$ as shown in Figure~\ref{Fig:case2}. Note there must exist some unmatched vertex $q$ that is adjacent to $u$ because after augmenting by the path $\{v_3^x = o_i, u, v_2^x, t_i\}$ the matched edge $(v_2^y = v_3^x, u)$ must be changed to $(v_2^y,v_1^y)$,  which can be done with an augmenting path of length three. After the augmentation step, we have $\phi(t_i) \geq \phi(q)$ (A), because there existed an augmenting path from $v_2^y$ to $q$. After $v_2^y$ is matched, assume without loss of generality that $q = HUN(u)$. By Lemma~\ref{lem:hun}, after $(v_2^y,u)$ is changed to $(v_2^y, v_1^y)$ we have $\phi(q=HUN(u)) \geq \phi(HUN(v_1^y)) \geq \phi(f_y)$ (B). Combining (A) and (B) we obtain  $\phi(t_i) \geq \phi(f_y)$.

Case 2: If $f_x$ charged the corresponding vertex of $v_3^x = v_2^y$,  then $f_y$ must charge the corresponding vertex of $v_3^y = v_2^x$. We will show that the corresponding vertex of $v_3^y$ is at least as heavy as $f_y$. Suppose that the corresponding vertex is strictly lighter than $f_y$ which is true if it is a terminus, say $t_i$ in the $i$th augmenting step.  By Lemma~\ref{lem:v3} we have  (1) at the step when the vertex  $v_3^y$ is matched but prior to matching it, the vertex $v_2^y$ is matched to some $u$, with $u \neq v_1^y$, such that  $\{v_2^y, v_3^y, u\}$ is a cycle; and (2) the augmenting path   is $\{v_3^y=o_i, u, v_2^y, t_i\}$. By symmetry and using the same argument as in Case 1 we get $\phi(t_i) \geq \phi(f_x)$. Since by assumption we have $\phi(f_x) \geq \phi(f_y)$, it follows that $\phi(t_i) \geq \phi(f_y)$. 

Note that each matched vertex has a unique corresponding vertex, since once they (the vertex and its corresponding vertex) are matched they will not be unmatched. So, to charge a vertex twice, a vertex $u$ must be considered by two failures (and  the corresponding vertex of $u$ must be charged  twice). But two failures cannot consider the same vertex. For two failures in different alternating paths, it is not possible since the alternating paths are vertex disjoint.  
For two failures in the same alternating path, by our charging method,  no two failures consider the same vertex for  charging purposes. 
$\quad \qed$

\begin{theorem} 
\label{thm:mainTheorem}
Algorithm~\ref{Approx} computes a ${2}/{3}$-approximation for the MVM problem.
\end{theorem}

{\bf Proof}:
Let $M_A$ be the matching computed by the approximation algorithm, and $M_{opt}$ be a matching of
maximum vertex weight. Consider all paths in the symmetric difference between $M_A$ and $M_{opt}$.  
Let $\phi(F)$ denote the sum of weights of all the failures,   let $\phi(M_{opt})$ denote the weight of the maximum-weighted matching, and let  $\phi(M_A)$ denote the weight of the approximate matching. 
Then, $\phi(M_{opt}) = \phi(M_A) + \phi(F) - \phi(M_A \setminus M_{opt}) \leq \phi(M_A) + \phi(F)$, 
and we know from Lemma~\ref{lem:compensating-vertices} that $\phi(F) \leq \frac{1}{2}\phi(M_A)$ since for every failure we have two distinct  vertices that are at least as heavy as the failures. 
Hence 
$\phi(M_{opt}) - \phi(M_A)  \leq  \phi(F) \leq \frac{1}{2}\phi(M_A)$.    Thus we have $\phi(M_{opt}) \leq \frac{3}{2} \phi(M_{A})$. 
This completes the proof. 
$\quad \qed$

%% file: sectionExperiments.tex
\section{Experiments and Results}
\label{sec:sectionExperiments}

\subsection{Experimental Setup and Algorithms Being Compared}

\begin{table}[!htb]
 {
 \footnotesize
  \caption{The set of test problems.}
 \begin{tabular}[!h]{|l|rrrr|r|}
 \hline
 \textbf{Graph} &$|V|$ &\multicolumn{3}{c|}{Degree}&  $|E|$ \\
                &        & \textbf{Max.} & \textbf{Mean} & \textbf{SD/Mean}
                & \\
 \hline \hline
G34  & 2,000 	&4    &	4.00 & 0.00 &		4,000 \\
G39  & 2,000 &	210 & 11.78	&1.17 &	11,778 \\
de2010 & 24,115 &	45 & 4.81 &0.62&	58,028 \\
shipsec8 & 114,919 &	131 &	56.90 &0.25& 3,269,240 \\
kron\_g500-logn17 & 131,072 & 29,935 &	94.78 &4.40& 5,113,985 \\
mt2010 &	132,288 & 139 &	4.83 &0.74& 319,334 \\
fe\_ocean & 143,437 & 6 &	5.71 &0.12&	409,593 \\
tn2010 &  240,116 &	89 &	4.97 &0.60&	596,983 \\
kron\_g500-logn19 &	524,288 &	80,674 &	 106.46 &5.76&	21,780,787 \\
tx2010 &	914,231 &	121 &	4.87 &0.63& 2,228,136 \\
kron\_g500-logn21 &	2,097,152 &	213,904 &	 117.92 &7.47&	91,040,932 \\
M6 & 3,501,776 &	10 &	5.99 &0.14&	 10,501,936 \\
hugetric-00010 &	6,592,765 &	3 &	2.99 &0.01&	9,885,854 \\
rgg\_n\_2\_23\_s0 &  8,388,608 &	40 &	15.14 &0.26&	63,501,393 \\
hugetrace-00010 & 12,057,441 &	3 &	2.99 &0.01&	18,082,179 \\
nlpkkt200 &    16,240,000 &	27 &	26.60 &0.09&	215,992,816 \\
hugebubbles-00010 &    19,458,087 &	3 &	2.99 &0.01&	29,179,764 \\
road\_usa &    23,947,347 &	9 &	2.41 &0.39&	28,854,312 \\
europe\_osm &	50,912,018 &	13 &	2.12 &0.23&	54,054,660 \\
 \hline
 \end{tabular}
 
 \label{Problems}
 }
\end{table}

We used an Intel Xeon E5-2660 processor-based system (part of the Purdue University Community Cluster), called \emph{Rice}\footnote{\tiny https://www.rcac.purdue.edu/compute/rice/} for the experiments. The machine consists of two processors, each with ten cores running at 2.6 GHz (20 cores in total) with 25 MB unified L3 cache and 64 GB of memory. The operating system is Red Hat Enterprise Linux release 6.9. All code was developed using C++ and compiled using the g++ compiler (version: 4.4.7) using the -O3 flag. Our test set consists of nineteen real-world graphs taken from the University of Florida Matrix collection~\cite{FMC11} covering several application areas. Table~\ref{Problems} gives some statistics on our test set. The graphs are listed in increasing order of the number of vertices. The largest number of vertices of any graph is nearly 51 million, and the largest number of edges is nearly 216 million. For each graph we list the maximum and average vertex degrees and the ratio of the standard deviation of the degrees and the mean degree. The average degrees vary from $2$ to $118$, and the graphs are diverse with respect to their degree distributions. The three  {\tt kron\_g500} graphs of different sizes have
high maximum degrees, and high ratios of the standard deviation of the degrees and mean degree, but most problems have low values.

We compare the $2/3$-approximation algorithm for MVM (we will call this Two-thirds algorithm) with a number of other algorithms.  

The Exact algorithm for MVM is similar to Algorithm~\ref{Approx} except that there is no restriction on the augmenting path length, and it is discussed in Section~\ref{sec:sectionTwoThirdApprox}, and in 
more detail in~\cite{Dobrian+:VWM}.
The complexity of the Exact algorithm we have implemented is $O(nm)$.  The Spencer and Mayr algorithm~\cite{spencer1} has $O(m \sqrt{n}\, \log n)$ time 
complexity, but is more complicated to implement, it is not clear if it would lead to better practical performance, and  
our focus in this paper is on approximation algorithms for MVM with much lower  time complexity. 
We improved the practical performance of the Exact algorithm  for MVM by two modifications: 
(1) If a search for an augmenting path fails, we mark all visited vertices, and when these vertices  are encountered in a future search,  the algorithm quits searching along those paths. 
(2) At the step of matching a vertex $u_i$ we find the heaviest unmatched vertex $u_j$ in the sorted list of
5
 vertices such that $ j> i$. If an augmenting path from $u_i$ to a vertex $v$ is found such that  $\phi(v)=\phi(u_j)$,  then the algorithm stops the search and augments the matching. 

We have included an exact algorithm for the maximum edge-weighted matching problem (MEM) implemented in LEDA~\cite{LEDA, Mehlhorn+:Ledabook} in our comparisons. This is a primal-dual algorithm implemented with advanced priority queues and efficient dual weight updates, with time complexity $O(nm \log n)$~\cite{Mehlhorn+:matching}. Since this is a commercial code, we can only run the object code, and we ran it with no initialization and with a fractional matching initialization that obtains a $\{0, 1/2, 1\}$ solution to the linear programming formulation of maximum weighted matching by ignoring the odd-set constraints (computed combinatorially), and then rounding the solution to $\{0,1\}$ values~\cite{Applegate+:matching}. We call these two variants LEDA1 and LEDA2, respectively. 

The Greedy Half approximation algorithm for MVM  (Half)  matches the vertices  in non-increasing order of weights,  matching an unmatched vertex to a  heaviest unmatched neighbor, and then deletes other edges incident on the endpoints of the matched edge. Its time complexity is $O(m + n \log n)$~\cite{Dobrian+:VWM}. 

We used two implementations (Random and Round-Robin) of the $(2/3 - \epsilon)$ approximation algorithm for MEM  due to Pettie and Sanders~\cite{pettie2004simpler}, and Maue and Sanders~\cite{Maue+:matching} with $\epsilon =  0.01$. Before describing each implementation we will describe a 2-augmentation centered at a vertex $v$ which is an operation that is used in both 
implementations. We define an {\em  arm\/} of $v$ to be  either $\{v,u\}$ or $\{v,u,u'\}$, where $(v,u)$ is an unmatched edge, and $(u,u')$ is a matched edge. The {\em gain\/} of an augmentation or exchange of edges is the increase in weight obtained by the transformation. 
There are  two cases:\newline
Case 1) $v$ is unmatched: find an arm of $v$ with the highest positive gain.\newline
Case 2) $v$ is matched to a vertex $v'$: find the highest positive gain by checking the gains of the following paths or cycles:
\newline (1)  Alternating cycles of length four that include the edge $(v,v')$.
\newline (2) Alternating paths of length at most four, which is done as follows: Find two vertex disjoint arms of $v$, with the  highest gains $P$ and $P'$, then find an arm of $v'$ with highest gain $Q$.
If $P$ and $Q$ are vertex disjoint then $P \cup (v,v') \cup Q$ is a highest gain alternating path;  otherwise choose $P' \, \cup (v,v') \cup Q$ as a highest gain alternating path.
\newline
There are two implementations of this algorithm. 
The {\em Random implementation\/}  chooses a random vertex $v$ and performs a 2-augmentation centered at $v$ with the highest-gain.
This is repeated $k = \frac{1}{3} \log{\epsilon}^{-1}$ times.
The {\em Round-Robin implementation\/}  randomly permutes the order of vertices,  and for each vertex $v$ in the permuted order performs 2-augmentation with the highest-gain centered at $v$. This is  repeated for $k= \frac{1}{3} \, \log{\epsilon^{-1}}$ phases. If no further improvement can be achieved after finishing a phase then the algorithm quits.
The algorithm can be initialized with the $1/2$-approximation algorithm called the Global Paths algorithm (GPA)~\cite{Maue+:matching}, which 
sorts the edges in non-increasing order of their weights. It constructs sets of paths and  cycles {\em of even length} by considering the edges in non-increasing order of their weights. Then it computes
a maximum weight matching for each path and cycle by dynamic programming, and it deletes the matched edges and their adjacent edges. The algorithm repeats until all edges are deleted.
The time complexity of the GPA algorithm is $O(m \log n)$, and that of the Round-robin $2/3$-approximation algorithm is 
$O(m \log {\epsilon}^{-1})$. 
Maue and Sanders~\cite{Maue+:matching} have reported that the Round-robin implementation with GPA initialization computed heavier matchings than the other three variants albeit at the expense of higher running times; we have obtained similar results,  and find that the Round-robin implementation with no initialization was the fastest among the four variants. Hence we report results from these two variants, called RR and GPA-RR, respectively. 

The final algorithm we implemented is a  
$(1 - \epsilon)$-approximate scaling  algorithm for MEM  (Scaling)  due to Duan and Pettie~\cite{DuanP-approxMWM}, with the choice of $\epsilon = 1/3$, $1/4$, and $1/6$.
The algorithm is based on a primal dual formulation of the problem with relaxed feasibility and complementary slackness conditions imposed at each scale.
The time complexity of the  Scaling algorithm is $ O(m \, \epsilon^{-1} \log(\epsilon^{-1}))$. 

In total, we have two exact algorithms for MEM and MVM, and four approximation algorithms. The exact MEM algorithm and  the $(2/3-\epsilon)$-approximation algorithm have two options for  initialization.  

Integer weights of vertices were generated uniformly at random in the range $[1 \, 1000]$, and real-valued weights were chosen randomly in the range $[1.0, 1.3]$.  The reported results are average of ten trials of randomly generated weights. The standard deviations for run-time, weight ratio, and cardinality ratio are close to zero, so there is not much  variation on these metrics for each  algorithm.

\subsection{Performance of the Algorithms} 

\begin{sidewaystable}[!htb]
 {
 
 \footnotesize
 \caption{Running times (seconds) and relative performance  of several  exact and approximation algorithms for the MEM and MVM problems. 
 The exact algorithms include the LEDA implementations for MEM, with no initialization and a fractional matching initialization, and the exact MVM algorithm described in this paper. 
 The approximation algorithms include the 1/2-MVM, the 2/3-MVM,  Round Robin $2/3-\epsilon$ MEM with $\epsilon= 0.01$ with and without GPA initialization, and the $1-\epsilon$-Scaling MEM approximation algorithm with $\epsilon =1/3$. Vertex weights are random integers in the range $[1 \, 1000]$. The three groups of  problems indicate those for which the LEDA implementation with no initialization terminated in under four hours; those for which the LEDA implementation with a fractional matching initialization terminated in under four hours but the first algorithm did not; and a problem for which none of the exact algorithms terminated in 100 hours. }
 \label{Time1}
 \medskip
 \begin{tabular}[!h]{|l|r||r|r|r|r|r|r|r|}
 \hline
                 &   Time (s)   & \multicolumn{7}{c|}{Relative Performance}	\\
 \textbf{Graph} & {Exact} 
                & {Exact} 
                & {Exact}
                & {$1-\epsilon$}
                 & {GPA, RR}
                  & {RR}
                  & {2/3-} 
                & {1/2-}  
                \\
                & LEDA1   
                & LEDA2 
                & MVM 
                & Scal.
                & \multicolumn{2}{c|}{$2/3-\epsilon$} 
                & MVM 
                & MVM \\
                & 	&    &    & $\epsilon = 1/3$ &\multicolumn{2}{c|}{$\epsilon=0.01$} & &\\
 \hline \hline

G34	 & 	1.4E-1	&	1.9E+0	&	1.3E+1	&	1.9E+1	&	2.0E+1	&	4.0E+1	&	4.5E+2	&	7.4E+2	\\
G39	 & 	1.2E+0	&	1.0E+2	&	4.6E+1	&	9.4E+1	&	8.7E+1	&	1.5E+2	&	2.2E+3	&	5.7E+3	\\
de2010	 & 	3.9E+0	&	1.4E+1	&	7.6E+0	&	2.7E+1	&	3.7E+1	&	8.0E+1	&	7.3E+2	&	1.3E+3	\\
kron\_g500-logn17	 & 	1.2E+2	&	1.4E+1	&	3.2E+1	&	3.0E+1	&	1.5E+1	&	6.9E+1	&	1.6E+3	&	4.4E+3	\\
mt2010	 & 	1.5E+1	&	1.0E+1	&	1.1E+1	&	1.9E+1	&	2.2E+1	&	6.2E+1	&	4.3E+2	&	7.3E+2	\\
fe\_ocean	 & 	6.5E+2	&	7.8E-1	&	5.8E+0	&	8.7E+2	&	8.0E+2	&	2.0E+3	&	1.5E+4	&	2.5E+4	\\
tn2010	 & 	2.3E+2	&	2.3E+1	&	1.7E+1	&	9.2E+1	&	1.6E+2	&	4.3E+2	&	3.2E+3	&	5.3E+3	\\
kron\_g500-logn19	 & 	6.1E+2	&	1.5E+1	&	3.5E+1	&	2.2E+1	&	1.4E+1	&	7.6E+1	&	1.8E+3	&	4.8E+3	\\
tx2010	 & 	1.9E+3	&	2.9E+1	&	2.1E+1	&	1.5E+2	&	2.6E+2	&	7.2E+2	&	5.3E+3	&	1.0E+4	\\
kron\_g500-logn21	 & 	3.3E+3	&	1.6E+1	&	3.1E+1	&	1.8E+1	&	1.4E+1	&	6.2E+1	&	1.7E+3	&	5.2E+3	\\
road\_usa	 & 	1.2E+3	&	9.0E+0	&	1.8E+1	&	3.5E+0	&	6.8E+0	&	1.8E+1	&	1.2E+2	&	1.7E+2	\\
europe\_osm	 & 	4.1E+3	&	1.6E+1	&	4.9E+1	&	7.0E+0	&	1.1E+1	&	3.5E+1	&	2.1E+2	&	2.8E+2	\\[1ex]
\hline																	
Geo. Mean	 & 		&	1.2E+1	&	1.9E+1	&	3.5E+1	&	3.8E+1	&	1.1E+2	&	1.2E+3	&	2.4E+3	\\[1ex]
\hline		
	 & 		&	Time (s)	&	\multicolumn{6}{|c|}{Relative Performance} \\ 
shipsec8	 & 		&	5.9E+0	&	2.1E-1	&	3.3E+0	&	1.1E+0	&	4.7E+0	&	9.9E+0	&	1.5E+2	\\
M6	 & 		&	1.2E+3	&	6.5E-1	&	1.3E+1	&	3.1E+1	&	8.8E+1	&	5.7E+2	&	1.3E+3	\\
hugetric-00010	 & 		&	3.7E+2	&	4.1E-1	&	3.9E+0	&	8.1E+0	&	2.1E+1	&	1.4E+2	&	2.2E+2	\\
rgg\_n\_2\_23\_s0	 & 		&	6.0E+3	&	2.7E+0	&	3.7E+1	&	2.7E+1	&	8.6E+1	&	3.8E+2	&	1.7E+3	\\
hugetrace-00010	 & 		&	6.4E+2	&	4.4E-1	&	3.9E+0	&	7.2E+0	&	2.0E+1	&	1.3E+2	&	2.0E+2	\\
hugebubbles-00010	 & 		&	1.8E+3	&	5.7E-1	&	5.9E+0	&	1.2E+1	&	3.2E+1	&	2.2E+2	&	3.3E+2	\\[1ex]
\hline																	
Geo. Mean	 & 		&		&	5.8E-1	&	7.3E+0	&	9.3E+0	&	2.8E+1	&	1.4E+2	&	4.0E+2	\\[1ex]
\hline				
            &       &       &               & \multicolumn{5}{|c|}{Time(s)} \\
nlpkkt200	&		&		&		        &	263.6	&	786.7	&	185.5	&	51.9	&	8.2	
	\\
 \hline
 \end{tabular}
 }
 \end{sidewaystable}


In Table~\ref{Time1} we group the problems into  three sets based on our results. 
In the first set, the time taken by the Exact MEM algorithm from LEDA without initialization (LEDA1), and the relative performance of the other algorithms (the ratio of the time taken by LEDA1 to the time taken by the other algorithm), are reported. 
Numbers greater than one indicate that the latter algorithms are faster.
For the second set of problems, the LEDA algorithm with no initialization did not complete in four hours. Hence we report the time taken by  LEDA2, the  code with fractional matching initialization,
and relative performance for the other algorithms. 
For the third set consisting of one problem, none of the exact algorithms completed in  100 hours, and we report the run times of the approximation algorithms. 

On the first set of problems, in geometric mean, the exact algorithms  LEDA2 and MVM are $12$ and $19$ times, respectively,  faster than LEDA1;  the Scaling algorithm and the GPA-RR  algorithms are about $35$ and $38$ times faster, respectively; and  the RR algorithm  is $110$
times faster; the Two-thirds  MVM algorithm is $1,200$ times faster, and the Half algorithm is $2,400$ times faster. 

On the second set of problems, the Exact MVM algorithm is slower than the exact MEM algorithm LEDA2 by a factor of about $1.7$. 
The approximation algorithms are all faster than LEDA2, 
the fastest again being the Half algorithm (by a factor of $400$), and the Two-thirds algorithm is faster by a factor of $140$. 
The scaling and the RR algorithms  are $7$ and $28$ times faster than LEDA2. 


For the {\tt nlpkkt200} problem, the Two-thirds  algorithm computed the matching in $52$ seconds on the integer weights;
the Half approximation algorithm took about  $8$ seconds, while the Scaling algorithm solved the same problem in $264$ 
seconds. The GPA-RR and  RR algorithms took $787$ and $186$ seconds, respectively.  This graph has an interesting structure. It comes from a nonlinear programming problem (it is a symmetric Kuhn-Tucker-Karush matrix), which can be partitioned into two subsets of vertices $V_1$ and $V_2$; vertices in the set $V_1$ are connected to each other and to vertices in the set $V_2$, but the latter is an independent set of vertices, i.e., no edge joins a vertex in $V_2$ to another vertex in $V_2$. There are $8,240,000$ vertices in $V_1$ and $8,000,000$ vertices in $V_2$. This structure creates a large number of augmenting paths for the exact algorithms, and we conjecture this is why these algorithms do not terminate. 

We also report the maximum time taken by an algorithm over all problems  on which  it terminated. 
For LEDA1, it is $4,076$ seconds on the {\tt europe\_osm} problem;
for LEDA2, $5,952$ s on the {\tt rgg} problem; the Exact MVM algorithm needed $3,139$ s on the {\tt huge\_bubbles} problem. 
The Scaling algorithm took $579$ s on the {\tt europe\_osm} problem, and the Half algorithm took $15$ s on the same problem.  The problem {\tt  nlpkkt200} needed the most time for the 
other approximation algorithms: $787$ s for the GPA-RR,  $186$ s for the RR algorithm, and $52$ s for the Two-thirds algorithm. 

\begin{sidewaysfigure}[ht]
\centering
\includegraphics[scale=0.9]{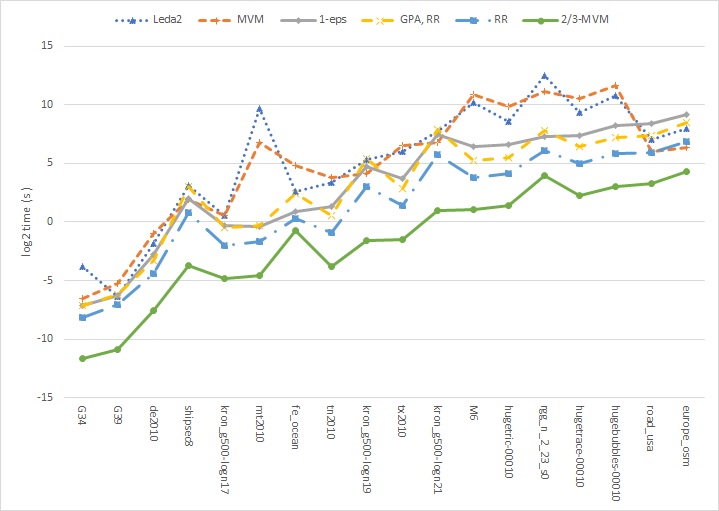}
\caption{Time taken by different algorithms (logarithmic scale), with  integer weights in $[1 \, 1000]$.}
\label{Fig:log2time}
\end{sidewaysfigure}

The runtimes of the some of these  algorithms  are plotted in a 
semi-logarithmic plot in Figure~\ref{Fig:log2time}.

%
\begin{table}[!htb]
 {
 \footnotesize
 \caption{The weights computed by the exact MEM and MVM algorithms, and the  gap to optimality of the weights of the matching obtained from the  approximation algorithms.   
For the last problem, the weights are shown since the 
Exact algorithms did not terminate. 
 Random integer weights in the range [1\ 1000] are used.}
 \label{Quality1}
 
 \begin{tabular}[!h]{|l|r||r|r|r|r|r|}
 \hline
                & Weight  & \multicolumn{5}{c|}{Gap to optimal weight $(\%)$} \\
 \textbf{Graph}  &  {Exact} 
                & {$1-\epsilon$-}
                & {GPA-RR}
                & {RR}
                 & {2/3-}  
                 & {1/2-} \\
                &algs.  & Scal.
                  & \multicolumn{2}{|c|}{$2/3-\epsilon$}
                   & MVM   &  MVM   \\
                 &     &$\epsilon=1/3$    
                 & \multicolumn{2}{c|}{$\epsilon=0.01$}  & & \\
 \hline \hline

G34	&	1.0E+6	&	1.89	&	0.31	&	0.30	&	0.47	&	2.88	\\
G39	&	1.0E+6	&	1.63	&	0.04	&	0.06	&	0.06	&	2.92	\\
de2010	&	1.2E+7	&	3.46	&	0.90	&	0.93	&	0.99	&	6.75	\\
shipsec8	&	5.7E+7	&	0.02	&	0.00	&	0.00	&	0.00	&	0.05	\\
kron\_g500-&        &           &           &           &           &   \\
logn17	&	4.4E+7	&	5.82	&	1.97	&	2.09	&	2.13	&	14.47	\\
mt2010	&	6.5E+7	&	3.86	&	1.10	&	1.14	&	1.21	&	7.61	\\
fe\_ocean	&	7.2E+7	&	1.06	&	0.14	&	0.15	&	0.21	&	2.26	\\
tn2010	&	1.2E+8	&	3.46	&	0.95	&	0.99	&	1.04	&	7.02	\\
kron\_g500-&         &          &           &           &           & \\
logn19	&	1.6E+8	&	5.29	&	1.72	&	1.81	&	1.95	&	14.33	\\
tx2010	&	4.5E+8	&	2.25	&	0.83	&	0.87	&	0.94	&	6.40	\\
kron\_g500-&        &           &            &          &           &  \\
logn21	&	5.5E+8	&	5.05	&	1.52	&	1.61	&	1.79	&	14.09	\\
M6	&	1.8E+9	&	0.70	&	0.16	&	0.16	&	0.21	&	2.39	\\
hugetric-&          &           &           &           &           & \\
00010	&	3.3E+9	&	1.57	&	0.50	&	0.64	&	0.81	&	4.43	\\
rgg\_n\_2\_23\_s0	&	4.2E+9	&	0.20	&	0.03	&	0.03	&	0.03	&	0.56	\\
hugetrace-&          &          &           &           &           & \\
00010	&	6.0E+9	&	1.56	&	0.49	&	0.62	&	0.79	&	4.39	\\
hugebubbles-&       &            &          &           &           & \\
00010	&	9.7E+9	&	1.57	&	0.50	&	0.63	&	0.81	&	4.42	\\
road\_usa	&	1.2E+10	&	3.35	&	1.16	&	1.50	&	1.74	&	7.81	\\
europe\_osm	&	2.5E+10	&	3.08	&	0.53	&	1.81	&	2.00	&	6.77	\\
\hline													
Geom. Mean	&		&	1.64	&	0.33	&	0.39	&	0.46	&	3.88	\\
\hline													
\\													
\hline	
            &       & \multicolumn{5}{|c|}{Weights} \\
nlpkkt200 	&		&	8.06E+09	&	8.07E+09	&	8.07E+09	&	8.08E+09	&	8.04E+09	
	
	\\
 \hline
 \end{tabular}
 
 }
\end{table}  

\begin{sidewaysfigure}[ht]
\centering
\includegraphics[scale=0.9]{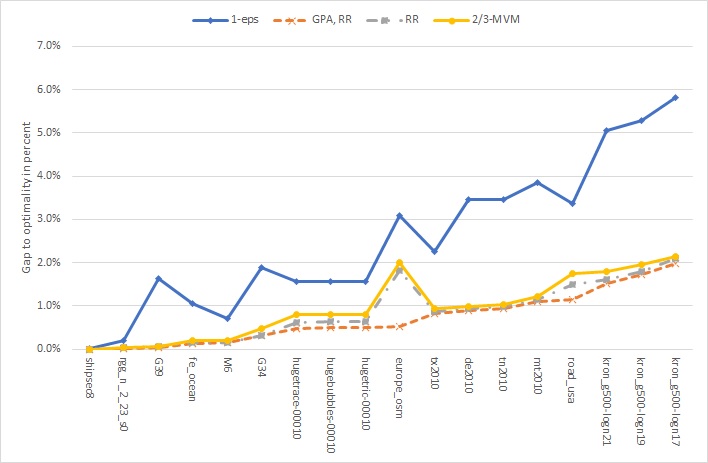}
\caption{Gaps to optimal weights for different algorithms, with integer weights in $[1 \, 1000]$. }
\label{Fig:GapToOptimality}
\end{sidewaysfigure}

We compare the weight of the matching computed by the  algorithms with integer weights in range $[1 \, 1000]$ in Table~\ref{Quality1}. 
All the exact algorithms compute the same maximum weight, which is reported in the first column; the approximation algorithms compute nearly optimal weights, and in order to differentiate among them, we report the gap to optimality as a percent. Hence we report $100 \,(1- w(A)/w_{\rm opt})$, where $w(A)$ is the weight computed by an  algorithm  $A$ and 
$w_{\rm opt}$ is the optimal weight computed by the exact algorithms. 
The Half algorithm computes weights higher than $96\%$ of the optimal, and the Scaling algorithm computes weights higher than $98\%$ of the optimal. The other approximation algorithms obtain weights higher than $99.99\%$ of the optimal. The best among these is  GPA-RR, which it accomplishes taking  run times at least  $20$ times higher than the Two-thirds algorithm.
Note that the weights obtained in practice are much better than 
the worst-case approximation guarantees. 
These results are plotted in Figure~\ref{Fig:GapToOptimality}. 

In Table~\ref{Time2}, we report run times  from the Exact MVM algorithm and the relative performance of the approximation algorithms when the vertex weights are real-valued in the range $[1 \ 1.3]$. These weights are favorable to the Scaling approximation algorithm, since the number of scales needed is low.  LEDA unfortunately does not work with real-valued  weights. In geometric mean, the Half algorithm is faster than the Exact MVM algorithm by a factor of $110$; the Two-thirds algorithm by a factor of $54$; and the other approximation algorithms are faster by factor less than $8$. 
On the {\tt nlpkkt200} problem, the Exact MVM algorithm did not terminate; notice that the Scaling algorithm is faster with the smaller range of weights here when compared to the integer weights with a larger range. The run times of the other approximation algorithms are consistent with the rankings discussed earlier.

Table~\ref{Quality2} includes results for the real-valued weights in the range $[1 \, 1.3]$.  The  Half approximation algorithm obtains about 89\% of the maximum weight (geometric mean of these problems), and is the worst performer. 
The other approximation algorithms are all comparable in the weights they compute, two or three percent off the optimal. Again the best performer is the GPA-RR  algorithm, which it achieves taking about a factor of nine more time than the Two-thirds algorithm.

\begin{table}[!htb]
 {
 \footnotesize
 \caption{The cardinality of the matchings obtained by the exact algorithms and the gap to optimality of the approximation algorithms.  
 For the last problem, cardinalities are shown since the Exact algorithm did not terminate. Random integer weights in [1 \, 1000].}
 \label{Cardinality1}
 
 \begin{tabular}[!h]{|l|r||r|r|r|r|r|}
 \hline
                 & Card.  & \multicolumn{5}{c|}{Gap to optimality ($\%$)} \\
\textbf{Graph}  &  {Exact} 
                & {$1-\epsilon$-}
                & {GPA, RR}
                & {RR}
                 & {2/3-}  
                 & {1/2-} \\
                & algs.  & Scal.
                  & \multicolumn{2}{|c|}{$2/3-\epsilon$}
                   & MVM   &  MVM   \\
                 &     &$\epsilon=1/3$    
                 & \multicolumn{2}{c|}{$\epsilon=0.01$}  & & \\
 \hline \hline
G34	&	 1,000 	&	7.59	&	2.53	&	2.53	&	3.58	&	9.01	\\
G39	&	 1,000 	&	7.29	&	1.03	&	1.13	&	1.22	&	9.74	\\
de2010	&	 11,853 	&	9.86	&	3.98	&	4.05	&	4.42	&	14.07	\\
shipsec8	&	 57,459 	&	0.91	&	0.14	&	0.15	&	0.17	&	1.13	\\
kron\_g500-&            &           &           &           &           & \\
logn17	&	 38,823 	&	6.82	&	2.90	&	3.06	&	2.85	&	16.93	\\
mt2010	&	 63,685 	&	9.96	&	4.08	&	4.15	&	4.56	&	14.52	\\
fe\_ocean	&	 71,718 	&	6.10	&	1.75	&	1.77	&	2.41	&	8.19	\\
tn2010	&	 117,989 	&	9.84	&	4.06	&	4.17	&	4.50	&	14.41	\\
kron\_g500&             &           &           &           &           &
\\
-logn19	&	 136,770 	&	5.65	&	2.40	&	2.51	&	2.45	&	16.24	\\
tx2010	&	 449,167 	&	7.35	&	3.73	&	3.81	&	4.26	&	13.55	\\
kron\_g500-&        &           &           &           &           &
\\
logn21	&	 482,339 	&	5.16	&	2.10	&	2.21	&	2.21	&	15.69	\\
M6	&	 1,750,888 	&	4.75	&	1.92	&	1.97	&	2.36	&	8.48	\\
hugetric-&       &           &           &           &           &
\\
00010	&	 3,296,382 	&	6.96	&	3.34	&	3.54	&	4.63	&	11.61	\\
rgg\_n\_2\_23\_s0	&	 4,194,303 	&	2.57	&	0.77	&	0.79	&	0.93	&	3.85	\\
hugetrace-&       &           &           &           &           &
\\
00010	&	 6,028,720 	&	6.88	&	3.26	&	3.45	&	4.55	&	11.51	\\
hugebubbles-&       &           &           &           &           &
\\
00010	&	 9,729,043 	&	6.94	&	3.32	&	3.52	&	4.61	&	11.58	\\
road\_usa	&	 11,325,669 	&	7.09	&	3.34	&	3.81	&	4.59	&	13.37	\\
europe\_osm	&	 25,149,787 	&	8.22	&	1.91	&	5.23	&	6.29	&	13.57	\\
\hline													
Geom. Mean	&		&	6.01	&	2.12	&	2.34	&	2.72	&	10.14	\\[1ex]
\hline	
            &       & \multicolumn{5}{|c|}{Cardinality} \\[1ex]
nlpkkt200 	&		&	7.88E+06	&	7.99E+06	&	7.99E+06	&	7.99E+06	&	7.82E+06	

	\\
 \hline
 \end{tabular}
 }
 \end{table}

Now we consider the cardinality of the matchings obtained by the algorithms in Tables~\ref{Cardinality1} and \ref{Cardinality2}. The exact algorithm for MVM  computes a  maximum cardinality
matching when the vertex weights are positive, and since the MEM problems are derived from MVM problems by summing the weights, the MEM algorithms also compute  maximum cardinality matchings. 
The Half approximation algorithm is about ten percent off the maximum cardinality, and the Scaling algorithm about six percent off. 
The other approximation algorithms are about two percent off the cardinality, with the GPA-RR algorithm the best performer. 
For eight of the nineteen problems, the exact algorithms obtained perfect matchings (cardinality equal to $n/2$ or $(n-1)/2$). 
Similar results are obtained when real weights with a smaller range is used, except that this time the Scaling algorithm finds higher cardinalities. 
To see how the Two-third algorithm fares against the Scaling algorithm  for a  smaller $\epsilon$, we compared it with $3/4$- and $5/6$-Scaling approximation algorithms.  
For integer weights in  the range $[1 \,1000]$ the Two-third  algorithm is $37$ times faster than the $3/4$-approximation, 
and  $54$ times faster than the  $5/6$-Scaling approximation algorithm. In geometric mean the Two-third algorithm obtained greater weight by $1.2\%$ and $1.9\%$, and higher cardinality by $3.1\%$ and $3.8\%$, relative to the $3/4$- and $5/6$-approximate Scaling algorithm. For real-valued weights in $[1  \, 1.3]$ the Two-third algorithm is $9.6$ and $12.6$ times faster than the $3/4$ and $5/6$-Scaling approximation algorithms, respectively. In geometric mean the Two-third  algorithm obtained greater weights by $4.7\%$ than the $3/4$-approximation;
it was worse by $0.9\%$ than the $5/6$-approximation; the cardinality was higher  by $4.7\%$ over the $3/4$-approximation, and worse by $1.1\%$ relative to the $5/6$-approximation.

%% file: sectionConclusions.tex
\section{Conclusions}
\label{sec:sectionConclusions}

We have described an augmentation-based $2/3$-approximation algorithm for 
MVM  on non-bipartite graphs whose time complexity is $O( m \log{\Delta}  + n \log n)$,
whereas the time complexity of an exact algorithm is $O(n^{1/2} m \log n)$. 
The approximation algorithm is derived in a natural manner from an  exact algorithm for computing maximum weighted matchings by restricting the length of augmenting paths to at most three.

The $2/3$-MVM algorithm has been implemented efficiently in C++, and on a set of nineteen graphs, some with hundreds of millions of  edges, it computes the approximate matchings in less than $52$ seconds. The weight of the approximate matching is greater than $98\%$ ($94\%$) of the weight of the Optimal matching for these problems on integer weights in  $[1 \, 1000]$ (real weights in $[1.0 \, 1.3]$).  
A Greedy Half-approximation algorithm is faster than the $2/3$-MVM algorithm by about a factor of two, but the weight it computes is lower, and can be as low as $84\%$ on the worst problem. All of these algorithms obtain  weights that are much higher than the worst-case approximation guarantees.

In addition, on geometric mean the $2/3$-MVM algorithm is faster than a Scaling based $(1-\epsilon)$ approximation algorithm by a factor of 28 on the integer weights in range $[1 \, 1000]$, which is expected due to the large overheads needed for the  handling of  blossoms and dual variable updates. While the Scaling algorithm is  faster on real-valued weights in a narrower  range $[1.0 \, 1.3]$ since there are  fewer scales, the $2/3$-MVM algorithm is still faster than it on average by  a factor of 7. The $2/3$-approximate MVM  algorithm obtains better matching weight than the Scaling approximation algorithm for relevant values of $\epsilon$ in all instances on the integer weights, and all but two graphs for the real-valued weights (on those two problems, the weights are close). 

The $(2/3 -\epsilon)$-approximation algorithm  for MEM, with Round-robin selection of augmentations and initialization with the Global Paths algorithm, computes higher weights than the $2/3$-approximation algorithm for MVM, but at a cost of an order of magnitude or more time. The weight differences are quite small for integer weights in a range $[1 \, 1000]$, but are about $0.7\%$ for real-valued weights in the range $[1 \, 1.3]$. 

We have also compared our algorithms with exact algorithms for the MEM problem from LEDA with a fractional matching initialization, and show that the exact MVM algorithm is quite competitive with it. 
The $2/3$-approximation algorithm for MVM is two to three  orders of magnitude faster than these exact algorithms, and there are problems on which the exact algorithms do not terminate in hundreds of hours. 

Half-approximation algorithms for MEM (e.g., the Locally Dominant edge 
and Suitor algorithms) do not require sorting and can be used or adapted to obtain $1/2$-approximate matchings for the MVM. 
The $2/3$-approximation algorithm for MVM designed here processes the vertices in non-increasing order of weights, but an algorithm based on the idea of searching for weight-increasing paths and cycles can avoid doing so, leading to a potentially parallel algorithm. 
This is the scope of our current work. 

\section*{Acknowledgements}

We thank Jens Maue (Zurich) and Peter Sanders of the Karlsruhe Institute of Technology for sharing the code for the GPA and $(2/3-\epsilon)$-approximation algorithms with us. 
\newline 
This work was supported in part by 
NSF grant  CCF-1637534; the 
U.S. Department of Energy through grant DE-FG02-13ER26135; and  the Exascale Computing Project (17-SC-20-SC), a collaborative effort of the DOE Office of Science and the NNSA.  

\FloatBarrier

%% file: Appendix.tex
\section{Tables with random real weights in the range $[1 \, 1.3]$}

\begin{table}[!htb]
 {
 
 \footnotesize
 \caption{Running times (seconds) of the  Exact algorithm, and relative performance of five algorithms: 
 the 1/2-MVM;  the 2/3-MVM;  Round Robin, GPA followed by Round Robin  $2/3-\epsilon$-approximation MEM for $\epsilon= 0.01$;  and the $1-\epsilon$-Scaling MEM approximation algorithm with $\epsilon =1/3$. Vertex weights are random reals in the range $[1 \, 1.3]$. The last row shows the times of the approximation algorithms for a problem on which the exact algorithm did not terminate.}
 \label{Time2}
 \medskip
 \begin{tabular}[!h]{|l|r|r|r|r|r|r|}
 \hline
                 &   Time (s)   & \multicolumn{5}{c|}{Relative Performance}	\\
 \textbf{Graph} & {Exact}
                & {$1-\epsilon$}
                 & {GPA, RR}
                  & {RR}
                  & {2/3-} 
                & {1/2-}  
                \\
                & MVM 
                & Scal.
                & \multicolumn{2}{c|}{$2/3-\epsilon$} 
                & MVM 
                & MVM \\
                &    & $\epsilon = 1/3$ &\multicolumn{2}{c|}{$\epsilon=0.01$} & &\\
 \hline \hline

G34	&	8.8E-3	&	6.4E+0	&	1.4E+0	&	2.8E+0	&	3.2E+1	&	4.1E+1	\\
G39	&	1.7E-2	&	7.0E+0	&	1.4E+0	&	2.5E+0	&	3.6E+1	&	7.2E+1	\\
de2010	&	2.3E-1	&	6.5E+0	&	2.3E+0	&	5.1E+0	&	5.0E+1	&	7.2E+1	\\
shipsec8	&	6.5E+0	&	1.4E+1	&	1.2E+0	&	5.1E+0	&	1.1E+1	&	1.6E+2	\\
kron\_g500-&       &           &           &           &           &
\\
logn17	&	3.4E+0	&	3.6E+0	&	4.2E-1	&	2.0E+0	&	5.0E+1	&	1.2E+2	\\
mt2010	&	7.3E-1	&	3.5E+0	&	1.1E+0	&	2.9E+0	&	2.2E+1	&	3.3E+1	\\
fe\_ocean	&	6.1E+1	&	4.5E+2	&	7.5E+1	&	1.9E+2	&	1.4E+3	&	2.3E+3	\\
tn2010	&	4.9E+0	&	8.7E+0	&	3.3E+0	&	9.0E+0	&	7.1E+1	&	1.1E+2	\\
kron\_g500-&        &           &           &           &           &
\\
logn19	&	1.8E+1	&	2.4E+0	&	4.1E-1	&	2.2E+0	&	5.4E+1	&	1.3E+2	\\
tx2010	&	3.4E+1	&	1.1E+1	&	4.5E+0	&	1.3E+1	&	1.0E+2	&	1.8E+2	\\
kron\_g500-&       &           &           &           &           &
\\
logn21	&	9.8E+1	&	2.0E+0	&	4.2E-1	&	1.9E+0	&	5.3E+1	&	1.6E+2	\\
M6	&	6.4E+2	&	3.4E+1	&	1.6E+1	&	4.7E+1	&	3.2E+2	&	7.0E+2	\\
hugetric-&        &           &           &           &           &
\\
00010	&	3.4E+2	&	1.7E+1	&	7.4E+0	&	2.0E+1	&	1.4E+2	&	2.0E+2	\\
rgg\_n\_2\_23\_s0	&	6.5E+2	&	1.6E+1	&	3.0E+0	&	9.1E+0	&	4.6E+1	&	1.8E+2	\\
hugetrace-&       &           &           &           &           &
\\
00010	&	5.7E+2	&	1.7E+1	&	6.5E+0	&	1.7E+1	&	1.2E+2	&	1.8E+2	\\
hugebubbles-&       &           &           &           &           &
\\
00010	&	1.2E+3	&	1.8E+1	&	7.9E+0	&	2.1E+1	&	1.5E+2	&	2.2E+2	\\
road\_usa	&	4.8E+1	&	5.8E-1	&	2.9E-1	&	7.7E-1	&	5.2E+0	&	7.2E+0	\\
europe\_osm	&	7.2E+1	&	5.0E-1	&	2.0E-1	&	6.4E-1	&	3.8E+0	&	5.1E+0	\\[1ex]
\hline													
Geom. Mean	&		&	7.7E+0	&	2.1E+0	&	6.1E+0	&	5.4E+1	&	1.1E+2	\\[1ex]
\hline													
            &       &\multicolumn{5}{|c|}{Time (s)} \\[1ex]
nlpkkt200 	&		&	4.35E+01	&	8.37E+02	&	1.90E+02	&	4.61E+01	&	8.03E+00	
	
	\\
 \hline
 \end{tabular}
 }
 \end{table}

\begin{table}[!htb]
 {
 \footnotesize
 \caption{The weight obtained by the exact MEM and MVM algorithms, and the  gaps to optimality of the matching computed by the  approximation algorithms.
For the last problem, weights are shown since the 
exact algorithm did not terminate. 
 Random real-valued weights in the range $[1\, 1.3]$ are used.}
 \label{Quality2}
 
 \begin{tabular}[!h]{|l|r||r|r|r|r|r|}
 \hline
                & Weight  & \multicolumn{5}{c|}{Gap to exact weight $(\%)$} \\
 \textbf{Graph}  &  {Exact} 
                & {$1-\epsilon$-}
                & {GPA, RR}
                & {RR}
                 & {2/3-}  
                 & {1/2-} \\
                & algs. & Scal.
                  & \multicolumn{2}{|c|}{$2/3-\epsilon$}
                   & MVM   &  MVM   \\
                 &     &$\epsilon=1/3$    
                 & \multicolumn{2}{c|}{$\epsilon=0.01$}  & & \\
 \hline \hline

G34	&	2.30E+03	&	1.89	&	2.27	&	2.23	&	3.17	&	8.16	\\
G39	&	2.30E+03	&	1.90	&	0.99	&	1.22	&	1.05	&	8.78	\\
de2010	&	2.73E+04	&	6.88	&	3.54	&	3.65	&	4.07	&	12.64	\\
shipsec8	&	1.32E+05	&	0.00	&	0.13	&	0.13	&	0.14	&	0.99	\\
kron\_g500-&            &           &           &           &           &
\\
logn17	&	9.08E+04	&	5.71	&	2.72	&	2.87	&	2.74	&	16.33	\\
mt2010	&	1.47E+05	&	6.29	&	3.69	&	3.77	&	4.09	&	13.27	\\
fe\_ocean	&	1.65E+05	&	1.44	&	1.55	&	1.56	&	2.13	&	7.23	\\
tn2010	&	2.72E+05	&	6.43	&	3.66	&	3.74	&	4.05	&	13.07	\\
kron\_g500-&       &           &           &           &           &
\\
logn19	&	3.20E+05	&	5.92	&	2.31	&	2.43	&	2.38	&	15.76	\\
tx2010	&	1.03E+06	&	3.71	&	3.34	&	3.44	&	3.82	&	12.28	\\
kron\_g500-&       &           &           &           &           &
\\
logn21	&	1.13E+06	&	5.51	&	2.02	&	2.13	&	2.15	&	15.26	\\
M6	&	4.03E+06	&	0.82	&	1.70	&	1.74	&	2.08	&	7.50	\\
hugetric-&       &           &           &           &           &
\\
00010	&	7.58E+06	&	2.73	&	2.97	&	3.17	&	4.13	&	10.30	\\
rgg\_n\_2\_23\_s0	&	9.65E+06	&	0.07	&	0.68	&	0.70	&	0.81	&	3.34	\\
hugetrace-&       &           &           &           &           &
\\
00010	&	1.39E+07	&	2.65	&	2.91	&	3.09	&	4.06	&	10.20	\\
hugebubbles-&       &           &           &           &           &
\\
00010	&	2.24E+07	&	2.76	&	2.95	&	3.15	&	4.11	&	10.26	\\
road\_usa	&	2.62E+07	&	6.62	&	3.05	&	3.51	&	4.21	&	12.25	\\
europe\_osm	&	5.79E+07	&	6.33	&	1.73	&	4.79	&	5.73	&	12.15	\\
\hline													
Geom. Mean	&		&	3.79	&	2.35	&	2.64	&	3.06	&	10.63	\\[1ex]
\hline	
            &       & \multicolumn{5}{|c|}{Weights} \\
nlpkkt200 	&		&	1.84E+7	&	1.84E+7	&	1.84E+7	&	1.84E+7	&	1.81E+7	
	\\
 \hline
 \end{tabular}
 
 }
\end{table}  
 \begin{table}[!htb]
 {
 \footnotesize
 \caption{The The cardinality of the matchings obtained by the exact algorithms and the gap to optimality of the approximation algorithms.  
 For the last problem, cardinalities are shown since the Exact algorithm did not terminate.  Random real weights in $[1 \, 1.3]$.}
 \label{Cardinality2}
 
 \begin{tabular}[!h]{|l|r||r|r|r|r|r|}
 \hline
                 & Card.  & \multicolumn{5}{c|}{Gap to optimality ($\%$)} \\
\textbf{Graph}  &  {Exact} 
                & {$1-\epsilon$-}
                & {GPA, RR}
                & {RR}
                 & {2/3-}  
                 & {1/2-} \\
                & algs.  & Scal.
                  & \multicolumn{2}{|c|}{$2/3-\epsilon$}
                   & MVM   &  MVM   \\
                 &     &$\epsilon=1/3$    
                 & \multicolumn{2}{c|}{$\epsilon=0.01$}  & & \\
 \hline \hline
G34	&	 1,000 	&	2.08	&	2.57	&	2.52	&	3.58	&	9.25	\\
G39	&	 1,000 	&	2.04	&	1.13	&	1.39	&	1.20	&	9.76	\\
de2010	&	 11,853 	&	7.26	&	3.94	&	4.06	&	4.53	&	13.96	\\
shipsec8	&	 57,459 	&	0.00	&	0.15	&	0.15	&	0.17	&	1.15	\\
kron\_g500-&       &           &           &           &           &
\\
logn17	&	 38,823 	&	4.99	&	2.86	&	3.01	&	2.85	&	16.92	\\
mt2010	&	 63,685 	&	6.54	&	4.09	&	4.18	&	4.53	&	14.56	\\
fe\_ocean	&	 71,718 	&	1.58	&	1.76	&	1.77	&	2.41	&	8.20	\\
tn2010	&	 117,989 	&	6.75	&	4.07	&	4.16	&	4.50	&	14.41	\\
kron\_g500-&       &           &           &           &           &
\\
logn19	&	 136,770 	&	5.17	&	2.41	&	2.53	&	2.46	&	16.24	\\
tx2010	&	 449,167 	&	3.84	&	3.73	&	3.82	&	4.26	&	13.57	\\
kron\_g500-&       &           &           &           &           &
\\
logn21	&	 482,339 	&	4.70	&	2.10	&	2.22	&	2.21	&	15.67	\\
M6	&	 1,750,888 	&	0.90	&	1.93	&	1.98	&	2.36	&	8.49	\\
hugetric-&       &           &           &           &           &
\\
00010	&	 3,296,382 	&	2.91	&	3.34	&	3.55	&	4.63	&	11.62	\\
rgg\_n\_2\_23\_s0	&	 4,194,303 	&	0.08	&	0.78	&	0.80	&	0.93	&	3.85	\\
hugetrace-&       &           &           &           &           &
\\
00010	&	 6,028,720 	&	2.83	&	3.27	&	3.46	&	4.55	&	11.51	\\
hugebubbles-&       &           &           &           &           &
\\
00010	&	 9,729,043 	&	2.95	&	3.32	&	3.52	&	4.61	&	11.58	\\
road\_usa	&	 11,325,669 	&	6.73	&	3.35	&	3.82	&	4.59	&	13.38	\\
europe\_osm	&	 25,149,787 	&	6.67	&	1.91	&	5.24	&	6.29	&	13.57	\\
\hline													
Geom. Mean	&		&	3.81	&	2.60	&	2.91	&	3.38	&	11.63	\\[1ex]
\hline	
            &       & \multicolumn{5}{|c|}{Cardinality} \\[1ex]
nlpkkt200 	&		&	8.00E+06	&	7.99E+06	&	7.99E+06	&	7.99E+06	&	7.82E+06	
	\\
 \hline
 \end{tabular}
 }
 \end{table}

%% file: main.bbl
\begin{thebibliography}{10}

\bibitem{LEDA}
Anonymous.
\newblock {LEDA 6.5 Description}.
\newblock Available from World Wide Web:
  \url{http://www.algorithmic-solutions.com/leda/index.htm}.
\newblock Accessed: 10/22/2018.

\bibitem{Applegate+:matching}
David Applegate and William Cook.
\newblock Solving large scale matching problems.
\newblock In {\em DIMACS Series in Discrete Mathematics and Theoretical
  Computer Science}, volume~12, pages 557--576. American Mathematical Society,
  1993.

\bibitem{avis1983survey}
David Avis.
\newblock A survey of heuristics for the weighted matching problem.
\newblock {\em Networks}, 13(4):475--493, 1983.

\bibitem{bell1}
Colin~E. Bell.
\newblock {Weighted matching with vertex weights: An application to scheduling
  training sessions in NASA space shuttle cockpit simulators}.
\newblock {\em European Journal of Operational Research}, 73(3):443--449, March
  1994.

\bibitem{coleman3}
Thomas~F. Coleman and Alex Pothen.
\newblock {The null space problem II. Algorithms}.
\newblock {\em SIAM J. Algebraic Discrete Methods}, 8(4):544--563, 1987.

\bibitem{FMC11}
T.~A. Davis and Y.~Hu.
\newblock {The University of Florida Sparse Matrix Collection}.
\newblock {\em ACM Transactions on Mathematical Software}, 38(1):1:1--1:25,
  2011.

\bibitem{Dobrian+:VWM}
Florin Dobrian, Mahantesh Halappanavar, Alex Pothen, and Ahmed Al-Herz.
\newblock A 2/3-approximation algorithm for vertex-weighted matching in
  bipartite graphs.
\newblock {\em SIAM Journal on Scientific Computing}, Oct. 2018.
\newblock {Accepted for publication. Available at arXiv:1804.08016.}

\bibitem{drake2003improved}
Doratha~E Drake and Stefan Hougardy.
\newblock Improved linear time approximation algorithms for weighted matchings.
\newblock In {\em Approximation, Randomization, and Combinatorial Optimization:
  Algorithms and Techniques}, pages 14--23. Springer, 2003.
\newblock (Lecture Notes in Computer Science, Volume 2129).

\bibitem{DuanP10b}
R.~Duan and S.~Pettie.
\newblock Approximating maximum weight matching in near-linear time.
\newblock In {\em Proceedings 51st IEEE Symposium on Foundations of Computer
  Science (FOCS)}, pages 673--682, 2010.

\bibitem{DuanP-approxMWM}
R.~Duan and S.~Pettie.
\newblock Linear time approximation for maximum weight matching.
\newblock {\em J. ACM}, 61(1), 2014.
\newblock Article 1.

\bibitem{duff1}
Iain~S. Duff and Jacko Koster.
\newblock On algorithms for permuting large entries to the diagonal of a sparse
  matrix.
\newblock {\em SIAM J. Matrix Anal. Appl.}, 22(4):973--996, 2000.

\bibitem{duff2}
Iain~S Duff and Bora U{\c{c}}ar.
\newblock Combinatorial problems in solving linear systems.
\newblock In Uwe Naumann and Olaf Schenk, editors, {\em Combinatorial
  Scientific Computing}, pages 21--68. 2009.

\bibitem{Hanke2010}
Sven Hanke and Stefan Hougardy.
\newblock New approximation algorithms for the weighted matching problem.
\newblock Research Report 101010, 2010, Research Institute for Discrete
  Mathematics, University of Bonn, 2010.

\bibitem{karypis1}
George Karypis and Vipin Kumar.
\newblock Analysis of multilevel graph partitioning.
\newblock In {\em Proceedings of the 1995 ACM/IEEE conference on
  Supercomputing}, page~29. ACM, 1995.

\bibitem{Khan1}
Arif Khan, Alex Pothen, Mostofa Patwary, Nadathur Satish, Narayanan Sunderam,
  Fredrik Manne, Mahantesh Halappanavar, and Pradeep Dubey.
\newblock Efficient approximation algorithms for weighted {$b$-Matching}.
\newblock {\em SIAM J. Scientific Computing}, 38:{S593--S619}, 2016.

\bibitem{manne2014new}
Fredrik Manne and Mahantesh Halappanavar.
\newblock New effective multithreaded matching algorithms.
\newblock In {\em IEEE 28th International Parallel and Distributed Processing
  Symposium}, pages 519--528. IEEE, 2014.

\bibitem{Maue+:matching}
Jens Maue and Peter Sanders.
\newblock Engineering algorithms for approximate weighted matching.
\newblock In {\em International Workshop on Experimental and Efficient
  Algorithms}, pages 242--255. Springer Verlag.
\newblock Lecture Notes in Computer Science, Vol.~4525.

\bibitem{Mehlhorn+:Ledabook}
Kurt Mehlhorn, Stefan Naher, and Stefan N{\"a}her.
\newblock {\em LEDA: a platform for combinatorial and geometric computing}.
\newblock Cambridge university press, 1999.

\bibitem{Mehlhorn+:matching}
Kurt Mehlhorn and Guido Sch\"{a}efer.
\newblock Implementation of {$O(nm \log n)$} algorithms for matchings in
  general graphs: the power of data structures.
\newblock {\em ACM Journal of Experimental Algorithmics}, 7:4--23, 2002.

\bibitem{Mendelsohn+:theorem}
N.~S. Mendelsohn and A.~L. Dulmage.
\newblock Some generalizations of the problem of distinct representatives.
\newblock {\em Canadian Journal of Mathematics}, 10:230--241, 1958.

\bibitem{pettie2004simpler}
Seth Pettie and Peter Sanders.
\newblock A simpler linear time 2/3- $\varepsilon$ approximation for maximum
  weight matching.
\newblock {\em Information Processing Letters}, 91(6):271--276, 2004.

\bibitem{pinar1}
Ali Pinar, Edmond Chow, and Alex Pothen.
\newblock Combinatorial algorithms for computing column space bases that have
  sparse inverses.
\newblock {\em Electronic Transactions on Numerical Analysis}, 22:122--145,
  2006.

\bibitem{pothen3}
Alex Pothen and Chin-Ju Fan.
\newblock Computing the block triangular form of a sparse matrix.
\newblock {\em ACM Trans. Math. Softw.}, 16(4):303--324, 1990.

\bibitem{preis1999linear}
Robert Preis.
\newblock Linear time 1/2-approximation algorithm for maximum weighted matching
  in general graphs.
\newblock In {\em STACS 99}, pages 259--269. Springer, 1999.

\bibitem{Schrijver:book}
Alexander Schrijver.
\newblock {\em {Combinatorial Optimization: Polyhedra and Efficiency. Volume A:
  Paths, Flows and Matchings}}.
\newblock Springer, 2003.

\bibitem{spencer1}
Thomas~H. Spencer and Ernst~W. Mayr.
\newblock Node weighted matching.
\newblock In {\em Proceedings of the 11th Colloquium on Automata, Languages and
  Programming}, pages 454--464, London, UK, 1984. Springer-Verlag.

\bibitem{tabatabaee1}
Vahid Tabatabaee, Leonidas Georgiadis, and Leandros Tassiulas.
\newblock {QoS provisioning and tracking fluid policies in input queueing
  switches}.
\newblock {\em IEEE/ACM Trans. Netw.}, 9(5):605--617, 2001.

\end{thebibliography}
